\documentclass[numberedappendix]{emulateapj}
\usepackage{natbib}
\usepackage{graphicx}
\usepackage{multirow}
\def\sun{\ifmmode\odot\else$\odot$\fi}
\def\micron{\hbox{~$\mu$m}}

\newcommand{\Neii}{\hbox{[Ne\,{\sc ii}]12.81\micron}}
\newcommand{\Neiii}{\hbox{[Ne\,{\sc iii}]15.56\micron}}
\newcommand{\Neva}{\hbox{[Ne\,{\sc v}]14.32\micron}}
\newcommand{\Nevb}{\hbox{[Ne\,{\sc v}]24.32\micron}}
\newcommand{\Oiv}{\hbox{[O\,{\sc iv}]25.89\micron}}

\def\Spitzer{\textit{Spitzer}}

\shorttitle{Local Luminous Infrared Galaxies. II AGN Activity}
\shortauthors{Alonso-Herrero  et al.}

\begin{document}

\title{Local Luminous Infrared Galaxies. II. AGN Activity from
  Spitzer/IRS spectra}
%\footnote{This work is based on observations made with the Spitzer Space Telescope,
%which is operated by the Jet Propulsion Laboratory, California Institute of
%Technology under NASA contract 1407.}
\author{Almudena Alonso-Herrero\altaffilmark{1,2,4,5}, Miguel
  Pereira-Santaella\altaffilmark{1}, George H. Rieke\altaffilmark{2},
and Dimitra Rigopoulou\altaffilmark{3}} 
 \altaffiltext{$\star$}{This work is based on observations made with the Spitzer Space Telescope, which is operated by the Jet Propulsion Laboratory, California Institute of Technology under NASA contract 1407}
\altaffiltext{1}{Centro de Astrobiolog\'{\i}a, INTA-CSIC, E-28850
  Torrej\'on de Ardoz, Madrid, Spain} 
\altaffiltext{2}{Steward Observatory, University
of Arizona, Tucson, AZ 85721, USA}
\altaffiltext{3}{Astrophysics Department, University
of Oxford, Oxford OX1 3RH, UK}
\altaffiltext{4}{Associate Astronomer}
\altaffiltext{5}{Current address: Instituto de F\'{\i}sica de
  Cantabria, CSIC-Universidad de Cantabria, 39005 Santander, Spain}

\begin{abstract}
We  quantify
the active galactic nucleus (AGN) contribution to the mid-infrared (mid-IR) and the total 
infrared (IR, $8-1000\,\mu$m) emission in a complete volume-limited 
sample of 53 local luminous infrared galaxies (LIRGs,
$L_{\rm IR}=10^{11}-10^{12}\,{\rm L}_\odot$). We decompose
the Spitzer Infrared Spectrograph (IRS) low-resolution $5-38\,\mu$m
spectra of the LIRGs into AGN and starburst  components using clumpy
torus   
models and star-forming galaxy templates, respectively. We find that
50\% (25/50) of local LIRGs have an AGN component detected with this
method.   There is good
agreement between these AGN detections through mid-IR spectral 
decomposition and other AGN
indicators, such as the optical spectral class, mid-IR spectral 
features and X-ray properties. Taking all the AGN indicators together,
the AGN detection rate in the individual nuclei of LIRGs is $\sim 62\%$. 
The derived AGN bolometric luminosities are in the range $L_{\rm bol}{\rm
  (AGN)}=0.4 -50\times 10^{43}\,{\rm erg \,    
    s}^{-1}$. The
AGN bolometric contribution to the IR luminosities of the galaxies is
generally small, with 70\% of LIRGs having 
$L_{\rm bol}[{\rm AGN}]/L_{\rm IR}\le 0.05$. Only $\simeq 8\%$ of
local LIRGs have a significant  AGN bolometric 
contribution $L_{\rm bol}[{\rm AGN}]/L_{\rm IR}>0.25$.  
From the comparison of our
results with literature results of ultraluminous infrared galaxies 
($L_{\rm IR}=10^{12}-10^{13}\,{\rm L}_\odot$), we confirm that in the local 
universe the AGN
bolometric contribution to the IR luminosity increases with the IR
luminosity of the galaxy/system.  If we add  up the AGN bolometric
luminosities we find that AGNs only account for 
$5\%^{+8\%}_{-3\%}$ of the total IR luminosity produced by local LIRGs
(with and without AGN detections).  This proves that the bulk of
the IR luminosity of local LIRGs is due to star formation activity. 
Taking the newly determined
IR luminosity density of LIRGs in the local universe, we then estimate  an
AGN IR luminosity density of 
 $\Omega_{\rm IR }^{\rm AGN} = 3\times10^5 L_\odot \,{\rm
  Mpc}^{-3}$ in LIRGs.

\end{abstract}
% \keywords{general ---}

\section{Introduction}\label{s:intro}

Luminous infrared (IR) galaxies (LIRGs) are defined as having infrared
($8-1000\,\mu$m) luminosities in the range $L_{\rm
  IR}=10^{11}-10^{12}\,{\rm L}_\odot$. Although an active galactic
nucleus (AGN) may contribute, it is believed that the bulk of their IR
luminosity is produced by dust 
heated by intense star-forming activity \citep{Sanders96}. Using the
prescription of \cite{Kennicutt98} the IR luminosities of LIRGs imply
star formation rates in the range $17-170\,{\rm M}_\odot \,{\rm
  yr}^{-1}$ for a Salpeter IMF. 

The early studies of the AGN activity in LIRGs made use mostly of
the optical spectral range. The classical
studies of \cite{Veilleux95} and \cite{Kim95}
showed that the fraction of sources containing an AGN increases at
higher IR luminosities, although the relative contributions of star
formation  and
AGN to the bolometric luminosity of the system were not well
determined. Recently \cite{Yuan2010} introduced a new approach for
spectral classification of the \cite{Veilleux95} sample. They found
that $\sim 22\%$ of LIRGs are 
classified as Seyfert or LINER, $\sim 37\%$ are AGN/starburst (SB)
composites, 24\% are HII-like, and the rest are ambiguous. 
They also showed that a large
fraction of the LIRGs previously classified as LINERs are AGN/SB
composites. 

The AGN activity of local LIRGs has
also been addressed using IR observations. 
\cite{Goldader97a,
  Goldader97b} obtained near-IR spectroscopy of a large sample of
LIRGs and found no new Seyfert 1 AGNs. 
In the mid-infrared (mid-IR) \cite{Valiante2009} 
used spectroscopy obtained with the {\it Spitzer} Infrared Spectrograph (IRS) 
to estimate the AGN emission at 6\,\micron \, of a sample of 
IR bright galaxies, as an
ingredient for their backward evolution model for IR
surveys. Locally they found an increasing contribution of the AGN emission at
6\micron \, at higher IR luminosities. 
They did not observe this tendency at higher 
redshifts, however, when modelling the mid-IR and submillimeter number
counts. \cite{Petric2011}, as part of the 
The Great Observatories All-Sky LIRG Survey \citep[GOALS, 
see][]{Armus2009}, used a number of mid-IR spectral indicators 
to derive an AGN contribution of 12\% for the GOALS LIRGs. 
At higher IR luminosities, most ultraluminous IR galaxies (ULIRGs, $L_{\rm
  IR}=10^{12}-10^{13}\,{\rm L}_\odot$) are only dominated bolometrically by AGN
emission at $L_{\rm IR} \gtrsim 5\times 10^{12}\,{\rm L}_\odot$
\citep{Nardini2008, Nardini2010}.

%{\rotate
\begin{table*}
\tiny

\tablewidth{0pt}
\caption{The Volume-limited Sample of Local LIRGs\label{tbl_sample}}

\begin{tabular}{llcccccccc}

\hline
\hline

Galaxy Name & \textit{IRAS} Name &
$v_{\rm hel}$ & Dist. & $\log L_{\rm IR}$ &
Ref. & Spectral Class & Ref. \\
& & (km s$^{-1})$ & (Mpc) & (L$_\odot$) & IR & & Class \\
\hline
NGC~23             & IRAS~F00073+2538 & 4478 & 64.7 & 11.11 & A1 &
composite &  B1, B2\\
MCG~+12-02-001     & IRAS~F00506+7248 & 4722 & 68.3 & 11.48 & A1 & HII & B1 \\
%MCG+12-02-002$^?$  & ``               & ??? \\
NGC~633            & IRAS~F01341$-$3735 & 5128 & 74.2 & 10.64 & A2 & composite & B2 \\
ESO~297-G012       & ``                & 5183 & 75.0 & 11.06 & A2 & HII & B3 \\
UGC~1845           & IRAS~F02208+4744 & 4601 & 66.5 & 11.12 & A1 & composite & B1 \\
UGC~02982          & IRAS~F04097+0525 & 5355 & 77.6 & 11.23 & A1 & HII & B4 \\
NGC~1614$^*$           & IRAS~F04315$-$0840 & 4778 & 69.1 & 11.67 & A1 &
composite & B2 \\
CGCG~468-002-NED01 & IRAS~F05054+1718 & 5268 & 76.3 & 10.55 & A3 & [NeV] & B5, B6 \\
CGCG~468-002-NED02 & ``                & 4951 & 71.6 & 10.83 & A3 & \nodata & \nodata \\
UGC~3351           & IRAS~F05414+5840 & 4433 & 64.1 & 11.26 & A1 & Sy2 & B7 \\
NGC~2369           & IRAS~F07160$-$6215 & 3196 & 46.0 & 11.13 & A1 & composite & B8 \\
NGC~2388           & IRAS~F07256+3355 & 4078 & 58.9 & 11.23 & A1 & HII & B1 \\
%NGC2389$^?$       & ``  & \\
MCG~+02-20-003     & IRAS~F07329+1149 & 4908 & 71.0 & 11.11 & A1 & composite & B1 \\
%MCG+02-20-002$^*$ & ``  & 5101 & ?? & ?? & \citealt{Surace2004} \\
NGC~3110           & IRAS~F10015$-$0614 & 5014 & 72.6 & 11.29 & A1 & HII & B2 \\
NGC~3256           & IRAS~F10257$-$4339 & 2790 & 40.1 & 11.66 & A1 & HII & B9 \\
ESO~264-G057       & IRAS~F10567$-$4310 & 5141 & 74.4 & 11.05 & A1 & HII & B10 \\
NGC~3690           & IRAS~F11257+5850 & 3057 & 44.0 & 11.43 & A4 & Sy2 & B11 \\
IC~694             & ``                & 3098 & 44.6 & 11.58 & A4 & LINER & B11 \\
ESO~320-G030       & IRAS~F11506$-$3851 & 3038 & 43.7 & 11.22 & A1 & HII & B8 \\
MCG$-$02-33-098-W  & IRAS~F12596$-$1529 & 4714 & 69.1 & 10.96 & A3 & composite & B2 \\
MCG$-$02-33-098-E  & ``                & 4775 & 69.1 & 10.48 & A3 & HII & B2 \\
IC~860             & IRAS~F13126+2453 & 3859 & 55.7 & 11.12 & A1 &
no &  B1 \\
MCG$-$03-34-064 & IRAS~F13197$-$1627 & 5009 & 72.5 & 11.09 & A2 & HII, [NeV], Sy1h & B2, B5, B6, B12  \\
NGC~5135           & IRAS~F13229$-$2934 & 4074 & 58.8 & 11.27 &A1 &
Sy2, [NeV] & B2, B5, B6 \\
ESO~173-G015       & IRAS~13242$-$5713 & 2948 & 42.4 & 11.57 & A1 & \nodata & \nodata \\
IC~4280            & IRAS~F13301$-$2356 & 4843 & 70.1 & 11.01 & A1 & HII & B8 \\
UGC~08739          & IRAS~F13470+3530 & 5000 & 72.3 & 11.05 & A1 & \nodata & \nodata \\
ESO~221-IG010      & IRAS~F13478$-$4848 & 3084 & 44.4 & 10.92 & A1 & HII & B8 \\
NGC~5653           & IRAS~F14280+3126 & 3513 & 50.7 & 10.98 & A1 & HII & B2 \\
NGC~5734           & IRAS~F14423$-$2039 & 3998 & 57.7 & 11.00 & A2 & composite & B8 \\
NGC~5743           & IRAS~F14423$-$2042 & 4121 & 59.5 & 10.84 & A2 &
HII, [NeV] & B8, B5 \\
IC~4518W           & IRAS~F14544$-$4255 & 4720 & 68.2 & 11.15 & A3 &
Sy2, [NeV] & B10 \\
IC~4518E           & ``               & 4518 & 66.3 & 10.28 & A3 &
no & B8 \\
Zw~049.057         & IRAS~F15107+0724 & 3858 & 55.7 & 11.21 & A1 & composite & B13 \\
NGC~5936           & IRAS~F15276+1309 & 3977 & 57.4 & 11.00 & A1 &
HII, composite & B1, B2 \\
NGC~5990           & IRAS~F15437+0234 & 3756 & 54.2 & 10.98 & A1 & Sy2,
[NeV] & B2, B5, B6 \\
NGC~6156           & IRAS~16304$-$6030 & 3253 & 46.9 & 11.12 & A1 & [NeV]& B5, B6 \\
\nodata & IRAS~F17138$-$1017 & 5161 & 74.7 & 11.39 & A1 & composite/HII & B2 \\
\nodata & IRAS~17578$-$0400 & 3931 & 56.7 & 11.31 & A1 & \nodata & \nodata \\
IC~4686$^*$        & IRAS~F18093$-$5744 & 4948 & 71.6 & 10.98 & A3 & HII & B2 \\
IC~4687            & `` & 5105 & 73.9 & 11.30 & A3 & HII & B2 \\
IC~4734            & IRAS~F18341$-$5732 & 4623 & 66.8 & 11.27 & A1 & HII & B10 \\
NGC~6701           & IRAS~F18425+6036 & 3896 & 56.2 & 11.03 & A1 & composite & B1 \\
NGC~6921$^*$      & IRAS~20264+2533 & 4327 & 62.5 & 10.37 & A2 & no & B14\\
MCG~+04-48-002     &    ``            & 4199 & 60.6 & 10.96 & A2 & HII, [NeV] & B14, B5, B6 \\
NGC~7130           & IRAS~F21453$-$3511 & 4837 & 70.0 & 11.38 & A1 &
Sy2:, [NeV] & B2, B5, B6 \\
IC~5179            & IRAS~F22132$-$3705 & 3363 & 48.5 & 11.18 & A1 & HII & B2 \\
NGC~7469           & IRAS~F23007+0836 & 4840 & 70.0 & 11.64 & A1 &
Sy1, [NeV] & B1, B2, B5, B6 \\
NGC~7591           & IRAS~F23157+0618 & 4907 & 71.0 & 11.11 & A1 &
composite, Sy2: & B1, B2 \\
NGC~7679           & IRAS~23262+0314  & 5162 & 74.7 & 11.12 & A1 &
Sy2/Sy1, [NeV] & B2, B5, B6 \\
NGC~7769           & IRAS~F23485+1952 & 4158 & 60.0 & 10.90 & A2 & composite & B13\\
NGC~7770           & IRAS~F23488+1949 & 4128 & 59.6 & 10.80 & A3 & HII & B2 \\
NGC~7771           & ``               & 4277 & 61.8 & 11.33 & A3 & HII, composite & B1, B2 \\

\hline

\end{tabular}

Notes.--- The heliocentric velocities are derived from the
high-resolution \Spitzer/IRS spectra.  For interacting galaxies, the
IR luminosities are given for the 
individual galaxies and the following column indicates the reference
for this. The spectral class is the optical class and/or whether the
\Neva \, line is detected. In this column ``no''  indicates that no
classification was possible from the optical spectroscopic data.\\ 
$^*$Not observed with the IRS and the IRS observation of NGC~1614 is
not well centered at its nucleus. The heliocentric velocities are from
NED.\\
REFERENCES. --- (A1) RBGS \citealt{SandersRBGS};  (A2)
\citealt{Surace2004};  (A3) This work from MIPS 24\micron \, data;
(A4) \citealt{Charmandaris02}; (B1) \citealt{AAH09PMAS}; (B2)
\citealt{Yuan2010};  (B3) \citealt{Kewley2001}; (B4)
\citealt{Veilleux95}  (B5) This work; (B6) \citealt{Petric2011}; (B7)
\citealt{Baan98}; (B8) \citealt{Pereira2011a}; (B9) \citealt{Lipari2000}; (B10)
\citealt{Corbett2003}; (B11) \citealt{GarciaMarin06}; (B12)
\citealt{Lumsden2001};  (B13) \citealt{Parra2010}; (B14) \citealt{Masetti2006}. 
\end{table*}
%}

In this paper we decompose the {\it Spitzer Space Telescope}/IRS
spectra of a complete   
volume-limited sample of LIRGs into SB and AGN components. The
main goal is to derive the mid-IR and bolometric AGN contributions in
these systems. Section~2 presents the sample and the
observations, while Section~3 describes the analysis of the data. 
Section~4 gives the results of the spectral decomposition of the IRS data
into the AGN and SB components. Section~5 compares the AGN
mid-IR detections with other AGN indicators including mid-IR, optical
and X-ray indicators. Section~6 discusses the AGN detection rate and AGN
bolometric contribution as a function of the IR luminosity and compares
them with results for ULIRGs in the literature. Section~7 summarizes
our conclusions.
Throughout this paper we assume the following cosmology $H_0 = 70$ km
s$^{-1}$Mpc$^{-1}$, $\Omega_M = 0.3$ and $\Omega_{\Lambda} = 0.7$.

\section{The Sample and Observations}\label{s:observations}

\subsection{The Volume-limited Sample of Local LIRGs}
We use the volume-limited 
sample of LIRGs  defined by \cite{AAH06}.  
This sample was originally drawn from
the {\it IRAS} Revised Bright Galaxy
Sample \citep[RBGS, ][]{SandersRBGS}, which is a complete flux-limited
survey at 60\micron \,  with flux densities greater than 5.24\,Jy and Galactic
latitude \textbar b\textbar\ 
$> 5$ deg. The distance limit imposed by \cite{AAH06} 
was chosen to  allow for Pa$\alpha$ observations 
with the narrow-band F190N filter of {\it HST}/NICMOS. The sample here
has been completed to include all the {\it IRAS} sources in the RBGS 
with $\log (L_{\rm IR}/{\rm L_\odot}) \geq 11.05$ 
and $v_{\rm hel} =$ 2750 -- 5300 km s$^{-1}$. 
For the assumed cosmology
the distances are in the range $\simeq 40-78\,$Mpc, with a median
value of 65\,Mpc.

The sample is presented in Table~1; it contains 45
\textit{IRAS} systems.  Eight \textit{IRAS } systems in our
  sample contain multiple galaxies, that is, they
are interacting galaxies,
pairs of galaxies, or galaxies with companions. 
These can be readily identified in
Table~1 as having the same {\it IRAS} name.
We note, however, that
MCG$-$03-34-063, which is part of IRAS~F13197$-$1627 with
MCG$-$03-34-064
\citep{Surace2004}, is at
$6394\,{\rm km \, s}^{-1}$ (from NED) and thus does not meet the
distance criterion of our sample.
 Two galaxies, NGC~5743 and NGC~7769 (see Table~1), have IRS spectroscopy
 (see Section~2.2), but were not  originally
 included in the {\it IRAS} RBGS. However, both galaxies 
are in interaction with RBGS {\it IRAS}
 sources \citep[see][]{Surace2004}. NGC~5743 is paired with NGC~5734 and shows a
 diffuse H$\alpha$ extension toward NGC~5734
 \citep{Dopita2002}. NGC~7769 is part of the NGC~7771/NGC~7770 group,
 and there is evidence that NGC~7769 is undergoing a direct encounter
 with NGC~7771 \citep{Nordgren97}. Additionally, these two galaxies not
   included in the RBGS have  IR luminosities similar to the rest of
   the galaxies (see Table~1) and thus we included them in our sample.
The sample contains a total of 53 individual galaxies.

The IR luminosities of the galaxies in the range 8--1000\micron,
L$_{\rm IR}$, were  calculated as defined by \citet{Sanders96}. The
\textit{IRAS} flux densities were taken from \citet{SandersRBGS} and 
\citet{Surace2004}. The latter work used image reconstruction techniques
to resolve the IR emission from the individual galaxies of 
interacting systems detected by {\it IRAS}. 
For the groups of galaxies and interacting
galaxies not resolved by \textit{IRAS} we assumed that the $L_{\rm
  IR}$ fraction of each component is
proportional to the {\it Spitzer}/MIPS 24\micron\  flux density (see
Section~2.3) 
fraction.  In Table~1 we list the IR luminosities for the individual
galaxies in our sample, rather than for the 
systems in the case of pairs of galaxies and galaxies in groups.
Note that some of the $\log L_{\rm IR}$ values are below the imposed
limit either because the galaxy is member of a pair or because of the
revised values of the distances.

\subsection{IRS Spectroscopic Data}
All individual galaxies in our sample, except for
  three\footnote{IC~4686 and NGC~6921 were not observed, and 
  the slit was not properly centered on the nuclear region of 
NGC~1614.}, were observed by the \textit{Spitzer}\slash IRS
\citep{Houck2004} instrument. We retrieved IRS
 spectroscopic data for
our sample of LIRGs from the {\it Spitzer} 
archive. Fifteen {\it IRAS} systems (16 galaxies) 
in our sample were observed 
in mapping mode and one in staring mode as part of our own  guaranteed
time observation (GTO) programs P30577 and P40479 (PI:
G.~H. Rieke) and were discussed in 
detail by \cite{AAH09Arp299} and \cite{Pereira2010IRSmapping}. 
Two more galaxies 
are from various programs (IC~860 from P1096 and NGC~7469 from P14).
 The rest of the sample are part of the GOALS legacy program
\citep{Armus2009} and were 
observed in staring mode. Observations at 
low-resolution ($R\sim$60--120) with the Short-Low (SL)
and Long-Low (LL) modules and at high-resolution with the 
($R\sim600$) Short-High (SH) and Long-High (LH) 
modules were available for all the galaxies. 

For the staring and mapping data reduction we followed
\cite{Pereira2010lines} and \cite{Pereira2010IRSmapping}, respectively. 
Briefly, we started with the basic calibrated data (BCD). 
Bad pixels were corrected using the
IDL package \texttt{IRSCLEAN}\footnote{The \texttt{IRSCLEAN} package
  is available at
  http://irsa.ipac.caltech.edu/data/SPITZER/
docs/dataanalysistools/tools/irsclean/}. Then
we subtracted the sky emission. 
For the staring data we extracted the spectra using the
standard programs included in the \textit{Spitzer} IRS Custom
Extraction (SPICE) package provided by the \textit{Spitzer} Science
Center (SSC) and the point source calibration. For the mapping
observations we constructed the data cubes using CUBISM
\citep{Smith2007}. 
The nuclear spectra of the galaxies observed in mapping mode 
were extracted using a 13\farcs4$\times$13\farcs4 aperture. Since the
data cubes are calibrated as  
extended sources  we applied a wavelength dependent aperture
correction to the mapping data to obtain spectra
comparable to those observed in staring mode. To calculate this
aperture correction we used the mapping observations of standard stars
(HR~7341, HR~6606, HR~6688, and HR~2491).

\subsection{IRAC and MIPS Imaging Data}
We used \textit{Spitzer} imaging data obtained with the IRAC 
\citep{Fazio2004} and the MIPS \citep{Rieke2004} instruments, 
as part of the GOALS legacy
program \citep{Armus2009}. We retrieved the BCD from the
\textit{Spitzer} archive. The BCD 
processing includes corrections for the instrumental response (pixel
response linearization, etc.), flagging of cosmic rays and saturated
pixels, dark and flat fielding corrections, and flux calibration based
on standard stars (see the IRAC and MIPS instrument handbooks for
details). We combined the BCD images into mosaics using the MOsaicker
and Point source EXtractor (MOPEX) software provided by the SSC using
the standard parameters.

We obtained integrated photometry of our LIRGs using an elliptical
  aperture chosen to  encompass the emission of each galaxy. This ellipse
  was calculated by fitting the emission of the galaxy ($3\sigma$ over
  the background) in the IRAC $8\,\mu$m band. 
%This corresponds to the emission with a surface brightness larger
%than 0.06Mj sr−1 at 8 μm. 
In this work we only used the IRAC
5.8\,$\mu$m and MIPS 24\,$\mu$m images (see Section~4.2). 
For the IRAC images we applied the extended source aperture correction
(up to $25-35\%$ in the IRAC $5.8\,\mu$m band)
to the integrated fluxes (see the IRAC Instrument
Handbook) to account for 
the diffuse scattering of the emission across the IRAC focal plane.

\begin{table*}
\caption{AGN Fractional Contributions within the IRS
  Slit from the AGN+SB Decomposition.\label{tbl_AGNwithinIRSslit}} 
\begin{center}
\begin{tabular}{lccccccc}
\hline
\hline
Galaxy       & $C_{6\mu{\rm m}}^{\rm IRS}$[AGN] & 
$C_{20\mu{\rm m}}^{\rm IRS}$[AGN]  &
 $C_{24\mu{\rm m}}^{\rm IRS}$[AGN] &    
 $C_{30\mu{\rm m}}^{\rm IRS}$[AGN]\\   
\\
\hline
NGC~23             &0.16     &0.10     &0.06    &0.02\\
NGC~633            &0.15     &0.07     &0.03    &0.01\\
CGCG~468-002-NED01 &0.54     &0.83     &0.82    &0.79\\       
UGC~3351           &0.25     &0.15     &0.11    &0.06\\       
NGC~2369           &0.26     &0.16     &0.10    &0.04\\       
NGC~3690           &0.81     &0.70     &0.60    &0.47 \\       
MCG~$-$02-33-098-W &0.09     &0.42     &0.33    &0.23\\       
MCG~$-$03-34-064   &0.77     &0.93     &0.91    &0.85\\       
NGC~5135           &0.54     &0.26     &0.16    &0.09\\       
UGC~08739          &0.33     &0.14     &0.08    &0.03\\
NGC~5653           &0.12     &0.10     &0.05    &0.02\\       
NGC~5743           &0.39     &0.28     &0.18    &0.09\\       
IC~4518W           &0.53     &0.71     &0.69    &0.60\\       
NGC~5936           &0.30     &0.13     &0.08    &0.04\\       
NGC~5990           &0.33     &0.30     &0.21    &0.12\\       
NGC~6156           &0.49     &0.73     &0.65    &0.57\\       
IC~4687            &0.12     &0.09     &0.05    &0.02\\       
NGC~6701           &0.25     &0.15     &0.08    &0.04\\       
MCG~+04-48-002     &0.20     &0.40     &0.41    &0.34\\       
NGC~7130           &0.36     &0.26     &0.18    &0.11\\       
NGC~7469           &0.49     &0.52     &0.43    &0.34\\       
NGC~7679           &0.27     &0.33     &0.24    &0.15\\       
NGC~7769           &0.41     &0.29     &0.19    &0.13\\       
NGC~7770           &0.58     &0.40     &0.27    &0.15\\       
NGC~7771           &0.19     &0.09     &0.05    &0.02\\       

\hline
\end{tabular}
\end{center}
\end{table*}

\section{Analysis}

\subsection{AGN+SB Spectral Decomposition of IRS Spectra }
The main goal of this work is to estimate quantitatively the AGN
activity, both in terms of mid-IR detection and bolometric
contribution, in a complete sample of local LIRGs. In this section we 
fit the {\it Spitzer}/IRS SL+LL spectra with a combination of
SB and AGN templates. Our approach to estimate the AGN
contribution is similar to  other works in the literature \citep{Nardini2008,
  Nardini2010}. Basically these methods hinge on the 
close similarity of the mid-IR spectra of high metallicity starbursts
\citep[see][]{Brandl2006}, and the strong differences in the spectral
shape of  AGN and SB emissions in this spectral range. 
 This makes it possible to use templates to represent 
the SB and AGN activity (or a power-law continuum 
for the latter), and thereby make a spectral decomposition of the observed
mid-IR spectra of galaxies into SB and AGN components.
The main difference  between our method and that of Nardini et al. is that we
use the entire spectral range of the SL+LL 
IRS spectra, $\sim 5-38\micron$, while they only 
used the $5-8\,\mu$m spectral range. In related works both
\cite{Valiante2009} and \cite{Deo2009}
derived the AGN contribution at different mid-IR wavelengths
of samples of local LIRGs and Seyferts, respectively
by subtracting a scaled SB template from the  observed IRS spectra.
We will compare our
results with  theirs   in Section~4.1.

\begin{figure*}
\center
\vspace{-0.1cm}
\includegraphics[width=0.45\textwidth]{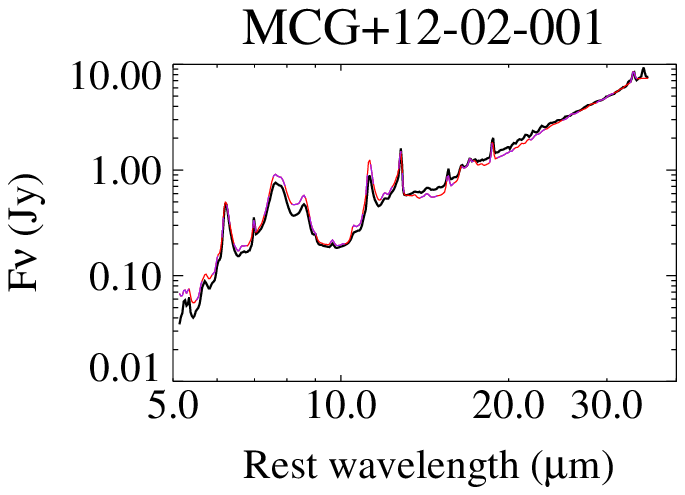}
\includegraphics[width=0.45\textwidth]{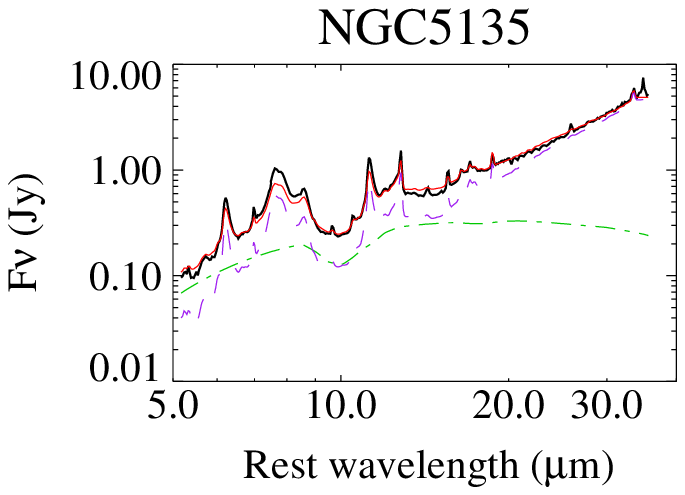}
\includegraphics[width=0.45\textwidth]{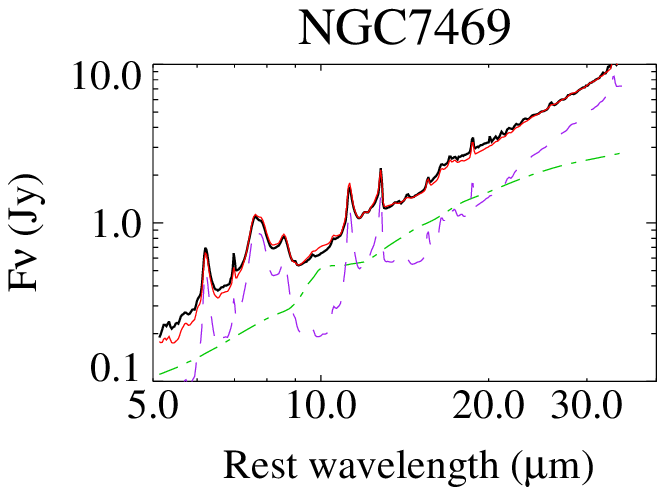}
\includegraphics[width=0.45\textwidth]{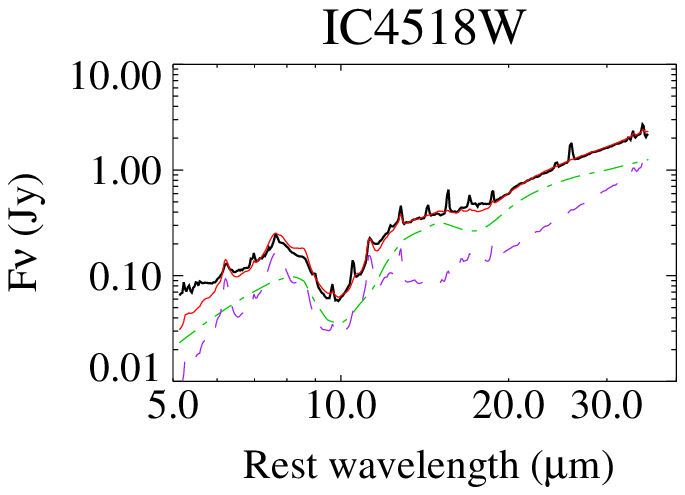}
\caption{Examples of the  AGN+SB best fits of the SL+LL IRS spectra for
four LIRGs in our sample with increasing AGN contributions from top
left to bottom right. The rest of the fits for the sample are
presented in Figures~A1 and A2. For reference, the best fit model of 
MCG~+12-02-001 does not require  AGN emission, whereas 
the AGN contribution within the {\it
  Spitzer}/IRS slit at $\lambda \ge 20\,\mu$m  increases from NGC~5135,
to NGC~7469, and IC~4518W (see Table~2).  
For each galaxy, the black line is the
observed spectrum in the rest-frame wavelength,
the dashed-dotted (green) line is the fitted torus model, the dashed
(purple) line is the scaled SB template, and the solid red line is the
sum of the fitted SB and AGN components. Note that for the AGN component the models only produce the continuum emission. Thus in the case of IC~4518W
the bright AGN emission lines are not reproduced by the sum of the AGN
and SB components. }
\label{fig_1}
\end{figure*}

The AGN emission in  
the mid-IR is emission reprocessed by warm dust in the putative 
dusty torus that surrounds the AGN \citep{Antonucci93}. Instead of
representing the 
AGN mid-IR  emission as a power law \citep[e.g., as done
by][]{Nardini2010} we used 
 the \cite{Nenkova2008a, Nenkova2008b}  clumpy dusty torus models 
together with  the BayesClumpy routine \citep[][see the Appendix
for more details]{BC} to fit the data. These torus
models (implemented in the \textit{CLUMPY} code) are found to reproduce well the
nuclear near and mid-IR  AGN emission of local Seyfert 
galaxies and PG quasars 
\citep{Nikutta2009, Mor2009, RA09, RA11, AAH11}. Moreover, clumpy dusty torus
models in general appear to reproduce better the observed IR emission of local
AGNs than smooth density torus models \citep[see e.g.][]{AAH03,
  Schartmann, Mullaney2011}. 
For the SB component we used the average spectrum of local
starbursts of \cite{Brandl2006}, as well as the LIRG
templates  of \cite{Rieke2009} in the
range $\log (L_{\rm IR}/{\rm L_\odot}) = 10.5 - 12$.

The \cite{Nenkova2008a, Nenkova2008b} clumpy torus models are defined
by six parameters describing the torus geometry and the
properties of the dusty clouds. In Appendix~A1 we give a short
description of the models. An additional parameter is the normalization 
needed to match the model to the observed IR data, which scales
directly with the AGN bolometric luminosity $L_{\rm bol} ({\rm AGN})$. 
Our use of these models 
depends on the scaling of the torus model and the AGN bolometric
luminosity, rather than on the individual torus parameters. \cite{AAH11}
recently showed that the values of 
$L_{\rm bol}(\rm AGN)$ derived from the modeling of the nuclear near
and mid-IR 
emission of local Seyfert galaxies with the \textit{CLUMPY } models 
were in good agreement with literature estimates.

Details of the AGN+SB decomposition method are given in Appendix~A2. 
Figure~1 shows a few examples of the results of the AGN+SB best fits
for four different LIRGs with increasing AGN
contributions. MCG~+12-02-001 shows  
 no evidence  for the presence of an AGN, while the mid-IR continuum
 emission of IC~4518W is dominated by the AGN component. NGC~5135 and
 NGC~7469 are intermediate cases. The AGN is clearly detected in 
NGC~5135, and
dominates the $\sim 6\,\mu$m emission within the IRS slit, but the SB
component accounts 
for most of the continuum emission at $\lambda > 15\,\mu$m. 
The AGN and SB components of NGC~7469 have similar contributions
within the IRS slit up
to $\sim 20\,\mu$m, but then at $> 20\,\mu$m the SB component
clearly takes over. The 
AGN+SB best fits for the rest of the sample are presented in
Figures~A1 and A2. For those galaxies requiring an AGN component to fit their
IRS spectra, Table~2 gives the AGN contribution within the {\it
  Spitzer}/IRS slit to the continuum emission
at four different reference wavelengths relatively free of features: 
6, 20, 24, and $30\,\mu$m.

\subsection{Line and Feature Measurements}

Most of the galaxies in our sample of LIRGs 
were part of the compilation of {\it Spitzer}/IRS
SH and LH fluxes of fine structure lines  of \cite{Pereira2010lines}.
For those not included there,  the emission line fluxes were measured
by fitting a Gaussian to the line plus a first order polynomial to the
local continuum \citep[see][for more
details]{Pereira2010lines}.  The lines were considered as 
detected if the peak of the emission line was  3$\sigma$ 
or move above the local continuum, where $\sigma$ is the standard
deviation of the local continuum.  The errors in the
  measured line fluxes depend on the signal-to-noise of the detected
  line, but are typically 10\% for the bright lines \Neii \, and
  \Neiii. For the fainter lines (\Oiv \, and  [Ne\,{\sc v}]) the
uncertainties are higher ($15-30\%$), except in bright AGN as these 
lines are intense and the uncertainties are approximately 10\%.
The \Oiv \, line fluxes (or upper limits) are
given in Table~3, together with those of the \Neii \, and \Neiii \,
emission lines.  
Additionally we detected the \Neva \, line in  five
galaxies in our sample 
not included in \cite{Pereira2010lines}: CGCG~468-002-NED01, NGC~5990,  
NGC~6156, MCG~+04-48-002, and NGC~7679 \citep[see
also][]{Petric2011}. Table~4 lists the  \Neva \, line detections
for our complete sample of LIRGs. 

\begin{figure}
\vspace{-0.1cm}
\center
\includegraphics[width=0.4\textwidth]{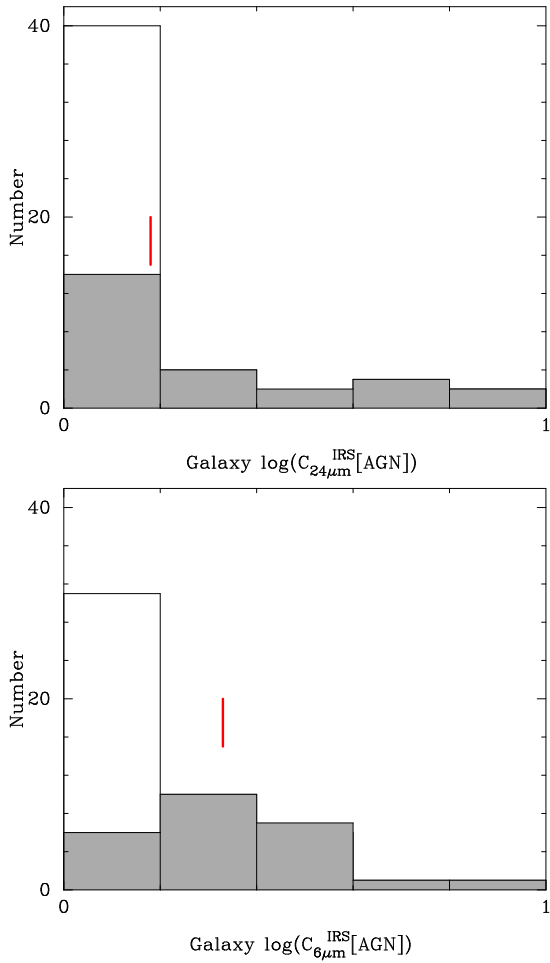}
\caption{Distributions of the AGN fractional contributions 
within the {\it Spitzer}/IRS
  slit of local LIRGs at $6\,\mu$m (lower panel) and $24\,\mu$m
  (upper panel). The AGN detections are shown as filled histograms, 
while the non-detections are
  shown as empty histograms. Also marked are the medians of the AGN
fractional  contribution (detections)
  distributions within the {\it Spitzer}/IRS slit  (see Table~5).}
\vspace{0.5cm}
\label{fig_2}
\end{figure}

To measure the apparent strength of the 9.7\micron \ silicate 
feature from the SL spectra we used the
definition of \cite{Spoon2007}, $S_{\rm Si} = \ln (f_{\rm obs}/f_{\rm cont})$ 
evaluated at 9.7\micron. To estimate the continuum emission 
 we used the same method as \cite{Pereira2010IRSmapping}, that
is, we fitted a power law between
the feature-free continuum pivots at 5.5\micron \, and 13\micron. 
We also measured the equivalent width (EW) of the 
6.2\micron \, aromatic  feature, also known as polycyclic aromatic
feature (PAH). The local continuum was measured using a linear
fit between 5.75\micron \ and 6.7\micron, and the feature was measured
between 5.9 and 6.5\micron. Table~4 reports the values of $S_{\rm Si}$ and
EW(6.2\micron \, PAH) for our sample of LIRGs. For objects with the
silicate feature in emission $S_{\rm Si}$ is positive, whereas it is
negative for objects with the feature in absorption.

\section{Results from the Decomposition of the {\it Spitzer}/IRS Spectra}

\subsection{AGN Dust Emission within the {\it Spitzer}/IRS Slits}
Using the AGN+SB spectral decomposition of the IRS low-resolution data we
detected an AGN component in half of the LIRGs in our sample (25/50)
with available data. Table~2 gives the AGN fractional
contribution within the slit at 6, 20, 24, and 30\micron \, for 
each of them. Figure~2 shows the distributions of the AGN  
fractional contributions within the IRS slit at two of the 
reference continuum wavelengths. 
On average the AGN fractional contribution within the IRS slits is
relatively small, as can be seen from Table~2. 
There are only six galaxies with a dominant AGN
contribution within the IRS slit at the four reference continuum wavelengths: 
CGCG~468-002-NED01, MCG~$-$03-34-064, IC~4518W, NGC~6156, NGC~7469, and
NGC~3690. 

For the sample of LIRGs the  median values of the AGN fractional contributions  
tend to be similar at 6 and $20\,\mu$m for our sample (median
$C_{6\mu{\rm  m}}^{\rm IRS}{\rm [AGN]}\simeq  
C_{20\mu{\rm m}}^{\rm IRS}{\rm [AGN]}= 0.3$, see Table~5), although
individual galaxies show larger variations. For the majority of local
LIRGs, the AGN fractional 
contributions decrease significantly at $\lambda > 20-30\,\mu$m, as  also
found in local Seyfert galaxies \citep{Deo2009} and X-ray selected AGNs
\citep{Mullaney2011}. The median 
value of the AGN fractional contribution at 24 and 30\micron \, for the LIRGs
with AGN detections are $C_{24\mu{\rm
    m}}^{\rm   IRS}{\rm [AGN]} = 0.18$ and $C_{30\mu{\rm
    m}}^{\rm   IRS}{\rm [AGN]}= 0.11$, respectively. For the four
Seyfert galaxies in our sample in common with 
\cite{Deo2009}, the AGN fractional contributions within the IRS slit 
at the common continuum wavelengths are in good agreement.

The fractional AGN contributions listed in Table~2 can also be used to
get a rough estimate of the sensitivity of our AGN+SB decomposition
method to the presence of a low-luminosity AGN. As discussed in 
the Appendix, we only considered an AGN detection when its
contribution was greater than $5-7\%$ at 20\micron. Therefore it
appears that our fitting method  can only
detect an AGN for fractional contributions $C_{6\mu{\rm
    m}}^{\rm IRS}{\rm [AGN]} \gtrsim 0.15$,  $C_{24\mu{\rm
    m}}^{\rm IRS}{\rm [AGN]} \gtrsim 0.05$ and  
 $C_{30\mu{\rm
    m}}^{\rm IRS}{\rm [AGN]} \gtrsim 0.02-0.04$.

\begin{table}
\caption{New IRS SH+LH Line Fluxes.\label{tbl_newIRSfluxes}}
%\begin{center}
\begin{tabular}{lccc}
\hline
\hline
Galaxy & [O\,{\sc iv}]	& [Ne\,{\sc ii}] & [Ne\,{\sc iii}] \\
%Galaxy & \Oiv	& \Neii & \Neiii \\
\\
\hline
ESO~297-G012	    &1.09	&46.68	&12.38\\
UGC~02982	    &1.58	&75.13	&9.07\\
CGCG~468-002-NED01   &8.98	&9.01	&7.47\\
CGCG~468-002-NED02   &$<1.07$	&43.32	&5.91\\
ESO~264-G057	    &0.91	&46.67	&3.52\\
ESO~173-G015	    &$<1.97$	&224.85	&27.34\\
IC~4280	            &$<0.18$	&50.76	&3.76\\
UGC~08739	    &$<0.77$	&28.85	&4.12\\
ESO~221-IG010	    &2.18	&86.74	&11.81\\
NGC~5990	            &11.77	&61.92	&17.86\\
NGC~6156	            &6.96	&35.47	&10.73\\
IRAS~17578$-$0400	    &$<0.98$	&56.29	&11.46\\
MCG~+04-48-002	    &8.42	&93.19	&20.00\\
NGC~7679	            &34.1	&77.29	&30.05\\
NGC~7770    	    &$<0.8$ 	&30.82	&7.82\\
\hline
\end{tabular}

Notes.--- Fluxes are in units of $10^{-17}\,{\rm W\, m}^{-2}$.
%\end{center}
\end{table}
\vspace{0.75cm}

\cite{Valiante2009} estimated the AGN emission at 6\micron \,
of a sample of LIRGs by subtracting a SB template from the observed
spectra in a much more limited
spectral range, $5.6-6.9\micron$. They employed the M82 spectrum
and the average spectrum of  
\cite{Brandl2006}, the latter  to account for the dispersion of
properties in star-forming galaxies. Both templates tend to have a
lower 6\micron \, continuum than the LIRG templates of \cite{Rieke2009}. 
A large fraction of the galaxies in our sample were included in their
work. For the AGN detections we find a relatively good agreement,
within $10-50\%$ of each other, for the AGN flux densities at
$6\,\mu$m. The largest discrepancy 
is for NGC~5135 for which we find a clear AGN component (see Figure~1)
at 6\micron, whereas the \cite{Valiante2009} upper limit is well
below our estimate.  Apart from our much broader
spectral range, the differences may also arise from 
the different SB templates.  We found a similar
behavior from our fits. In those cases where we obtained comparable
fits with the \cite{Brandl2006} average spectrum and the 
 LIRG $11.50{\rm L}_\odot$ template, the AGN fractional
contribution  within the
IRS slit at 6\micron \ using the latter template was
lower, almost by a factor of two, than using the former.
For those galaxies for which \cite{Valiante2009} did not
detect an AGN component, our estimated AGN 6\micron \, flux densities
are
compatible  with their upper limits.

%\tabletypesize{\scriptsize}
% \tabletypesize{\tiny}
\begin{table*}
\tiny

\caption{AGN Indicators\label{tbl_agn_indicators}}
\begin{tabular}{lccccccccc}
\hline
\hline
Galaxy Name &  Spectral class & [NeV] & $C_{6\mu{\rm
    m}}^{\rm tot}{\rm [AGN]}$ & $C_{24\mu{\rm
    m}}^{\rm tot}{\rm [AGN]}$ &  
 [OIV]/[NeII] & EW(6.2\micron\,PAH) & 
$S_{\rm Si}$\\
& &  Detection & \\
\hline
NGC~23          &    composite   & No  &    0.06&    0.05&    0.014& 0.52& -0.36\\
MCG~+12-02-001  &    HII         & No  & $<$0.05& $<$0.05&    0.016& 0.56& -0.33\\
NGC~633         &    composite   & No  &    0.08&    0.03&    0.037& 0.48& -0.30\\
ESO~297-G012    &    HII         & No  & $<$0.06& $<$0.05&    0.023& 0.61& -0.65\\
UGC~1845        &    composite   & No  & $<$0.07& $<$0.05&    0.029& 0.51& -0.94\\
UGC~02982       &    HII         & No  & $<$0.03& $<$0.04&    0.021& 0.57& -0.58\\
CGCG~468-002-NED01&  AGN         & Yes &    0.34&    0.80&    0.997& 0.10& -0.16\\
CGCG~468-002-NED02&  \nodata     & No  & $<$0.08& $<$0.05& $<$0.025& 0.37& -1.19\\
UGC~3351        &    Sy2         & No  &    0.07&    0.07&    0.012& 0.42& -1.16\\
NGC~2369        &    composite   & No  &    0.14&    0.08&    0.015& 0.41& -0.78\\
NGC~2388        &    HII         & No  & $<$0.08& $<$0.05&    0.012& 0.50& -0.51\\
MCG~+02-20-003  &    composite   & No  & \nodata & $<$0.05$^*$ & $<$0.019&
0.12& -1.47\\ 
NGC~3110        &    HII         & No  & $<$0.04& $<$0.04&    0.016& 0.57& -0.33\\
NGC~3256        &    HII         & No  & $<$0.05& $<$0.04& $<$0.026& 0.53& -0.45\\
ESO~264-G057    &    HII         & No  & $<$0.03& $<$0.04&    0.019& 0.51& -0.60\\
NGC~3690        &    Sy2         & No  &    0.44&    0.28&    0.143& 0.16& -0.60\\
IC~694          &    LINER       & No  & $<$0.04& $<$0.02& $<$0.064& 0.60& -1.80\\
ESO~320-G030    &    HII         & No  & $<$0.07& $<$0.04& $<$0.008& 0.51& -0.46\\
MCG~$-$02-33-098-W&    composite & No  &    0.05&    0.37& $<$0.017& 0.48& -0.02\\
MCG~$-$02-33-098-E&    HII       & No  & $<$0.08& $<$0.09&    0.039& 0.61& -0.37\\
IC~860          &    no          & No  & $<$0.05& $<$0.05& $<$0.173& 0.36& -1.58\\
MCG~$-$03-34-064  &    Sy1       & Yes &    0.64&    0.85&    2.054& 0.01& -0.19\\
NGC~5135        &    Sy2         & Yes &    0.39&    0.14&    0.658& 0.33& -0.37\\
ESO~173-G015    &    \nodata     & No  & $<$0.04& $<$0.04& $<$0.009& 0.33& -1.75\\
IC~4280         &    HII         & No  & $<$0.02& $<$0.03& $<$0.004& 0.52& -0.43\\
UGC~08739       &    \nodata     & No  & 0.12  & 0.06 & $<$0.027& 0.42& -1.18\\
ESO~221-IG010   &    HII         & No  &$<$0.02& $<$0.04&    0.025& 0.55& -0.32\\
NGC~5653        &    HII         & No  &    0.03&    0.04& $<$0.018& 0.49& -0.33\\
NGC~5734        &    composite   & No  & $<$0.03& $<$0.04&    0.012& 0.42& -0.45\\
NGC~5743        &    composite   & Yes &   0.14&    0.15&    0.065& 0.43& -0.57\\
IC~4518W        &    Sy2         & Yes &   0.34&    0.67&    2.119& 0.05& -1.22\\
IC~4518E        &    no          & No  &$<$0.04& $<$0.04&    0.019& 0.47& -0.56\\
Zw~049.057      &    composite   & No  & $<$0.06& $<$0.05& $<$0.029& 0.40& -1.09\\
NGC~5936        &    composite   & No  &    0.09&    0.07&    0.013& 0.50& -0.47\\
NGC~5990        &    Sy2         & Yes &    0.22&    0.18&    0.190& 0.13& -0.64\\
NGC~6156        &    AGN         & Yes &    0.05&    0.43&    0.196& 0.34&  0.57\\
IRAS~17138$-$1017 &    composite & No  & $<$0.07& $<$0.05& $<$0.028& 0.54& -0.66\\
IRAS~17578$-$0400 &    \nodata   & No  & $<$0.09& $<$0.04& $<$0.017& 0.52& -1.10\\
IC~4687         &    HII         & No  &    0.07&    0.05&    0.017& 0.63& -0.26\\
IC~4734         &    HII         & No  & $<$0.05& $<$0.05& $<$0.028& 0.40& -1.01\\
NGC~6701        &    composite   & No  &    0.09&    0.06&    0.029& 0.50& -0.56\\
MCG~+04-48-002  &    composite   & Yes &    0.09&    0.41&    0.090& 0.50& -0.92\\
NGC~7130        &    Sy2         & Yes &    0.20&    0.15&    0.186& 0.34& -0.30\\
IC~5179         &    HII         & No  & $<$0.04& $<$0.03&    0.017& 0.58& -0.28\\
NGC~7469        &    Sy1         & Yes &   0.32&    0.40&    0.125& 0.21& -0.13\\
NGC~7591        &    composite   & No  & $<$0.04& $<$0.04& $<$0.010& 0.38& -0.87\\
NGC~7679        &    Sy1/Sy2     & Yes &   0.12&    0.23&    0.441& 0.56& -0.05\\
NGC~7769        &    composite   & No  &    0.06&    0.10&    0.032& 0.41& -0.23\\
NGC~7770        &    HII         & No  &   0.29&    0.27& $<$0.026& 0.28& -0.33\\
NGC~7771        &    composite   & No  &   0.08&    0.04&    0.011& 0.45& -0.59\\
\hline
\end{tabular}

Notes.--- The spectral class is from Table~1. If a galaxy is classified 
as HII or composite, we list the composite classification. Those galaxies with
\Neva \, detections not classified as Seyferts from optical
spectroscopy are labeled as ``AGN''  or ``composite'' 
depending on whether the observed \Oiv/\Neii \, line ratio is high or
low, respectively
(see Section~5.1).  The [Ne\,{\sc v}] column
indicates if the
\Neva \,  high excitation
line is detected. $C_{6\mu{\rm  m}}^{\rm tot}{\rm [AGN]}$ and $C_{24\mu{\rm
    m}}^{\rm tot}{\rm [AGN]}$ are the AGN fractional contributions to
the total 6 and $24\,\mu$m emission respectively. [OIV]/[NeII]
is the observed line ratio, with typical uncertainties of $10-15\%$
for bright AGN, and  $15-30\%$ for the rest. The EW of the
$6.2\,\mu$m PAH features are measured in $\mu$m, and the typical
uncertainties are $0.05\,\mu$m. $S_{\rm Si}$ is the apparent 
strength of the 9.7\micron \, silicate feature and the typical
uncertainties are 0.02.
$^*$The integrated flux density is from {\it IRAS}
25\micron \, converted to MIPS $24\,\mu$m. 
\end{table*}

\subsection{AGN Contribution to the Total Mid-IR Continuum Emission}

The extent of the mid-IR  emission of LIRGs
shows no clear dependence on the 
IR luminosity of the system; some galaxies show relatively
compact emission ($\sim 1\,$kpc), while others show extended emission
over a few kpc \citep{Soifer2001, AAH06TReCS, TDS08,
  Pereira2010IRSmapping}. It is only at the high luminosity end of 
 LIRGs  ($\log (L_{\rm IR}/L_\odot) \gtrsim 11.80$) and in ULIRGs,
that the mid-IR emission starts appearing very compact 
\citep{Soifer2000, TDS10}. Moreover, the extent of the
emission may depend markedly on the wavelength of the emission. 
For instance, several studies found
that in LIRGs and star-forming galaxies the PAH emission tends to be more
extended than the mid-IR
continuum \citep[see e.g.][]{Helou2004, AAH09Arp299, Pereira2010IRSmapping}.

The IRS slits do not generally cover the full extent of the galaxies.
For the median distance of our LIRGs the $\sim 13\arcsec$ slits cover
approximately the central 4\,kpc. In this section we 
derive the AGN contribution to the total emission at two continuum
wavelengths. To do so, we measured
the total galaxy emission in the mid-IR
using  {\it Spitzer} imaging at two wavelengths 
IRAC 5.8\micron \, and MIPS 24\micron. The integrated flux densities
were then compared with the AGN emission at
these wavelengths from the AGN+SB decomposition. Table~4 lists
the AGN contribution to the 6 and 24\micron \, total emission, $C_{6\mu{\rm
    m}}^{\rm tot}{\rm [AGN]}$ and $C_{24\mu{\rm
    m}}^{\rm tot}{\rm [AGN]}$. For the AGN non-detections we give the upper
limits based on our estimated AGN detection limits as explained in
Section~4.1.  

The median AGN contribution to the total 6 and 24\micron
\, emission is 0.12 and 0.15, respectively, for those galaxies with an
AGN mid-IR detection (see Table~5). Only $\sim 6\%$ of local
LIRGs have their $24\,\mu$m emission dominated by the AGN. 
Unlike what we found for the AGN
fractional contributions within the IRS slits, for some LIRGs 
the AGN contribution to the total emission can be higher at 24\micron
\, than at 6\micron, especially for those LIRGs dominated by the AGN
contribution. It is possible
that the IRAC 5.8\micron \, maps include a significant contribution
from the 6.2\micron \, PAH feature \citep[see][and references
therein]{Helou2004}, as well as some photospheric emission. Additionally,
the emission from {\it CLUMPY } torus models is expected to be 
mostly isotropic at 24\micron, whereas that might  not be the case  at
6\micron \, for obscured views (i.e., type 2) of the AGN \cite[see
e.g., ][]{Nenkova2008b}.

In summary, only in 6\% (3/50) of local 
LIRGs is the mid-IR emission dominated by
the AGN ($C_{24\mu{\rm
    m}}^{\rm tot}{\rm [AGN]}>0.5$). About one-quarter of the sample 
(12/50) have intermediate AGN contributions ($C_{24\mu{\rm
    m}}^{\rm tot}{\rm [AGN]}=0.1-0.5$), while the remaining $\sim 70\%$
(35/50)  have little ($C_{24\mu{\rm
    m}}^{\rm tot}{\rm [AGN]}\simeq 0.04-0.1$) or no AGN contribution
in the mid-IR. 

%\subsection{Bolometric AGN luminosities and contribution to the IR
%  luminosity } 

\begin{table}
\caption{Statistics of the AGN Contribution in LIRGs with 
AGN Detections from Spectral Decomposition}
\begin{center}
\begin{tabular}{lccc}
\hline
\hline
Quantity & Mean& Median & $16^{\rm th}-84^{\rm th}$\\
\hline
  \multicolumn{4}{c}{Within IRS Slit}\\
\hline
$C_{6\mu{\rm m}}^{\rm IRS}{\rm [AGN]}$  & 0.36 & 0.33 & $0.16-0.54$\\
$C_{20\mu{\rm m}}^{\rm IRS}{\rm [AGN]}$ & 0.34 & 0.28 & $0.10-0.71$\\
$C_{24\mu{\rm m}}^{\rm IRS}{\rm [AGN]}$ & 0.28 & 0.18 & $0.06-0.65$\\
$C_{30\mu{\rm m}}^{\rm IRS}{\rm [AGN]}$ & 0.21 & 0.11 & $0.02-0.57$ \\
\hline
  \multicolumn{4}{c}{Total Emission}\\
\hline
$C_{6\mu{\rm m}}^{\rm tot}{\rm [AGN]}$  & 0.18 &0.12 & $0.07-0.34$\\
$C_{24\mu{\rm m}}^{\rm tot}{\rm [AGN]}$ & 0.24 &0.15 & $0.05-0.43$\\
\hline
  \multicolumn{4}{c}{AGN Bolometric Contribution}\\
\hline
$L_{\rm bol}{\rm [AGN]}/L_{\rm IR}$    & 0.12 &0.05 & $0.02-0.26$\\
\hline
\end{tabular}
\end{center}
Notes.--- The columns are the mean, median (50-th percentile), and the 
16-th and 84-th percentiles. The numbers in the last
column are the 68\% confidence interval. This 
would be equivalent  to $\pm 1\sigma$ of the 
distribution around the mean value in a normal distribution.\
\vspace{0.75cm}
\end{table}

\section{Comparison with other AGN Diagnostics}

\subsection{Spectroscopic Activity Class}
The nuclear activity (AGN versus star formation activity) 
of local LIRGs has been
the subject of numerous studies  using optical spectroscopy \citep[see among others,
][]{Veilleux95, Wu98, Kewley2001,   AAH09PMAS, Yuan2010}.
These works made use of the classical optical spectral diagnostic
diagrams \citep{Veilleux87} and/or the more recent
classification scheme based on observations of large numbers of
galaxies and modeling \citep{Kauffmann2003, Kewley2006}. 
Table~1 lists the spectral activity
class for the 53 galaxies in our sample and the relevant references. Out
of the 42 individual LIRGs with an 
optical class there are:  17 pure H\,{\sc ii}-like nuclei, 
15 composite\footnote{According to the \cite{Kewley2006} scheme, see
  also \cite{Yuan2010}.} (AGN/SB) nuclei, and 10
AGN (Seyfert 1, 
Seyfert 2, LINER). Two more LIRGs do not have an optical class but have
 \Neva \,  detections and high
\Oiv/\Neii \, ratios \citep[Table~4, and see
also][]{Petric2011}, and thus are clearly 
AGNs. Finally, NGC~5743 and MCG+04-48-002 were optically 
classified as HII, but both have a [Ne\,{\sc v}] detection. 
Both have also relatively low \Oiv/\Neii \, ratios, which are typical of
star-forming galaxies (see Table~4 and 
Section~5.4). Therefore, although these two galaxies
do contain an AGN, they are likely to be composite in nature. 
The rest (7 LIRG nuclei) do not have an optical
classification. 
In summary, 23\% of our sample are AGNs, 32\% are HII-like,  32\% are AGN/SB
composites, and the remaining 13\% do not have a spectral classification.

\begin{figure}
\center
\vspace{0.5cm}
\includegraphics[width=0.45\textwidth]{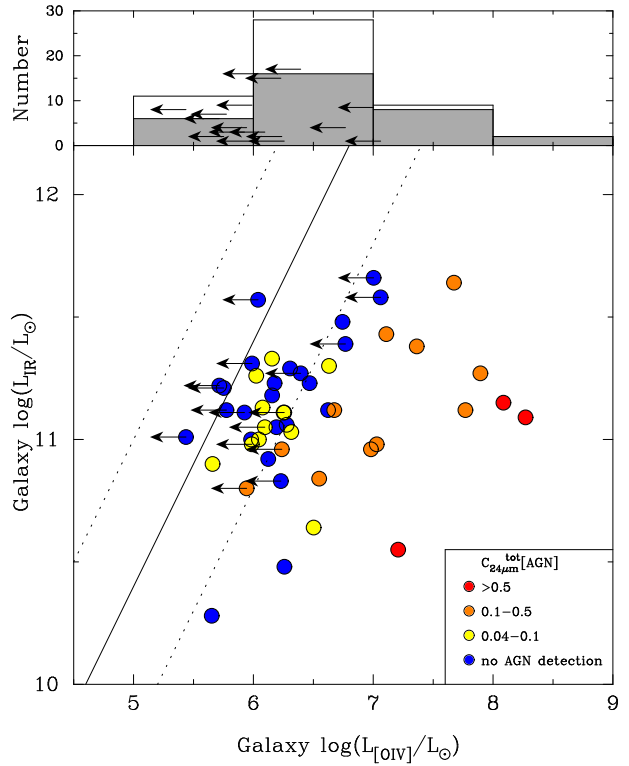}
\caption{{\it Upper panel.} 
Distribution of the [O\,{\sc iv}]$25.89\,\mu$m emission line
  luminosities (detections and upper limits) of our sample of local
  LIRGs. The filled histograms   correspond to those LIRGs with an AGN
  detection using optical and  mid-IR methods. The arrows mark the
  upper limits to  
  the [O\,{\sc iv}]$25.89\,\mu$m luminosity in each bin. 
{\it Lower panel.} Comparison between the observed \Oiv \,
luminosities and the IR luminosities of local LIRGs. The solid line is
the relation expected for
purely  star-forming galaxies derived by \cite{Pereira2010lines} (see text for 
details), while the dotted lines are the $\pm 1\sigma$ dispersion of
the relation. The LIRG observations are color coded according to the
AGN contribution to the integrated $24\,\mu$m emission:
$C_{24\mu{\rm
    m}}^{\rm tot}{\rm [AGN]}>0.5$, $C_{24\mu{\rm
    m}}^{\rm tot}{\rm [AGN]} = 0.1-0.5$, and $C_{24\mu{\rm
    m}}^{\rm tot}{\rm [AGN]} =0.04-0.1$, and the 
AGN non-detections, for which $C_{24\mu{\rm
    m}}^{\rm tot}{\rm [AGN]} \lesssim  0.02-0.05$.}
\vspace{0.2cm}
\label{fig_3}
\end{figure}

As can be seen from Table~4, all 13 LIRGs with an optical
classification as Seyfert nuclei and/or a \Neva \,
detection have an AGN component detected through the AGN+SB
decomposition of the IRS spectra. The AGN contributions at 24\micron
\, for these vary from 7\% for UGC~3351 to almost 90\% for MCG~$-$03-34-064.
Among the 14 optically composite nuclei with IRS spectra, 8 have an AGN
component detected with AGN contributions at 24\micron \, in the range
$4-28\%$. Three of the 15 pure HII nuclei have an AGN
component, but with a small AGN contribution at 24\micron \, except
for NGC~7770.

\subsection{AGN Bolometric Luminosities and Comparison with X-ray Estimates}
 The AGN bolometric luminosities derived from the torus model
fits (Section~3.1) 
are between 0.4 and $50\times 10^{43}\,{\rm erg \, s}^{-1}$, with
a median value of $L_{\rm bol}[{\rm AGN}] = 2\times 10^{43}\,{\rm erg
  \, s}^{-1}$. Galaxies classified as Seyferts
tend to show the highest values of $L_{\rm bol}{\rm [AGN]}$, whereas
the non-Seyferts have a median value of $L_{\rm bol}[{\rm AGN}]\sim
1\times 10^{43}\,{\rm erg   \, s}^{-1}$.  The AGN
bolometric contribution to the IR luminosity of the individual nuclei 
range between 1 and 70\%, with an average value of
$L_{\rm bol}{\rm [AGN]}/L_{\rm IR}=12\%$ (median of 5\%) for those LIRGs with
an AGN detection (see Table~5). For the AGN non-detections we estimate
that if an AGN is present its bolometric contribution would be 
$L_{\rm bol}{\rm [AGN]}/L_{\rm IR}<1\%$.

\cite{Pereira2011a} studied the X-ray emission of a representative, in
terms of $L_{\rm IR}$ and nuclear activity,
sample of local LIRGs drawn from the sample presented in this paper. 
Table~6 lists the eight Seyferts in common with their work. 
In this section we compare  our estimates of the AGN
bolometric luminosity for the Seyfert galaxies with those obtained 
by \cite{Pereira2011a} from the hard 
$2-10\,$keV luminosities, as well as with the values derived from
the luminosity of the 6.4\,keV FeK$\alpha$ line for the Compton-thick
sources. The AGN bolometric
luminosities were computed using  the bolometric corrections of
\cite{Marconi2004}. All these values are listed in Table~6. 
The last column in Table~6 gives the differences of the 
 two estimates (taken in logarithmic units) of the AGN bolometric
luminosity, which are on average $0.4\,$dex. This is similar to the
differences in bolometric luminosity estimates found by 
\cite{AAH11} for a sample of nearby Seyferts. We excluded NGC~7679
because it appears to be highly variable in X-rays \citep[see the
discussions by ][]{DellaCeca2001, Pereira2011a}.

From the upper limits of the luminosity of the hard X-ray FeK$\alpha$ emission
line, \cite{Pereira2011a} concluded that in the rest of their sample,
if there is an AGN the bolometric luminosity should be $\lesssim 10^{43}\,{\rm
erg \, s}^{-1}$. This is again consistent with the typical AGN bolometric
luminosities for the non-Seyfert LIRGs in our sample.

\begin{table}
\caption{Comparison between X-ray and AGN+SB Decomposition AGN Bolometric
  Luminosities }
\begin{tabular}{lcccc}
\hline
\hline
Galaxy        & $L_{\rm 2-10kev}$  & $L_{\rm bol}[{\rm AGN}]$  &
$L_{\rm bol}[{\rm AGN}]$ & Diff. \\
 & & X-ray & IRS fit \\
\hline
NGC~3690         & 3.9$^{\rm a}$ &55.  &88.  & -0.20\\
MCG~$-$03-34-064 & 15           &290. &330. & -0.06\\
NGC~5135         & 10$^{\rm a}$  &180. &53.  & 0.53\\
IC~4518W         & 2.2          &26.  &71.  & -0.44\\
MCG~+04-48-002   & 6.9          &120.  &21. & 0.76\\
NGC~7130         & 10$^{\rm a}$  &180. &40.  & 0.65\\
NGC~7469         & 17           &340. &460. & -0.13\\
NGC~7679         & 3.7/0.4$^{\rm b}$ &56./4. &75.  & \nodata \\
\hline
\end{tabular}

Notes.--- All the luminosities are in units of $10^{42}\,{\rm erg \,
  s}^{-1}$.  The $2-10\,$keV luminosities (corrected for
absorption) and the $L_{\rm bol}$[AGN] from X-rays are 
from \cite{Pereira2011a}. The typical uncertainties are $5-10$\% and 
25\%, respectively. The formal errors for 
$L_{\rm bol}$[AGN] from the IRS spectral decomposition are
$0.1-0.2\,$dex. The last column is the difference between the two
estimates of the AGN bolometric luminosity taken in logarithmic units.\\
$^{\rm a}$Estimated from the luminosity of the
FeK$\alpha$ line. 
$^{\rm b}$Variable? First value is from
\cite{DellaCeca2001} and second value from \cite{Pereira2011a}. 

\end{table}

%\begin{figure}
%\center
%\includegraphics[width=0.45\textwidth,angle=-90]{figure5.ps}
%\vspace{0.5cm}
%\label{fig_5}
%\end{figure}

\subsection{The [O\,{\sc iv}]$25.89\,\mu$m High Excitation Line}
AGNs can be readily identified in the mid-IR via the detection of
high-excitation emission  lines \citep[see e.g.][]{Genzel98, Sturm2002,
Melendez2008}, especially the \Neva \, and \Nevb \,  lines. These are 
unlikely to be produced by star formation, but they are not always
detected in relatively bright AGNs \citep[see][]{Weedman2005,
  Pereira2010lines}.  The detection rate of the \Neva \,
line in our complete sample of local LIRGs is $\sim 22\%$ (11/50),
as can be seen from Table~4.

\begin{figure}
\center
\includegraphics[width=0.45\textwidth,angle=-90]{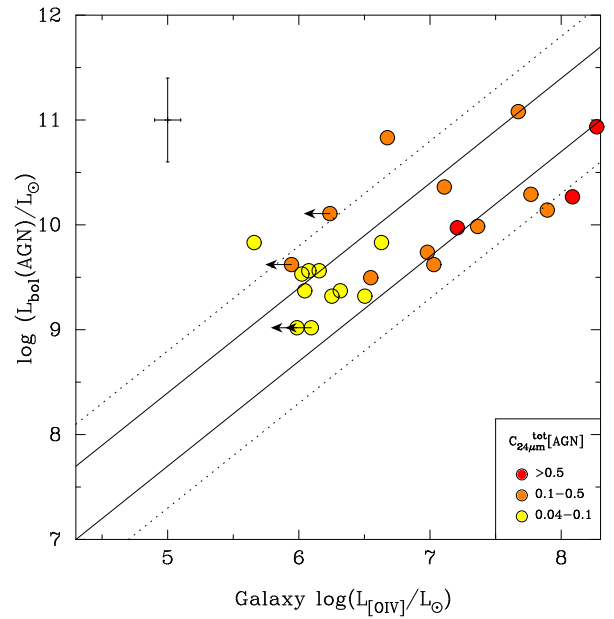}
\caption{AGN bolometric luminosity (from the AGN+SB fit to the IRS
  spectral) versus  the observed 
[O\,{\sc iv}]$25.89\,\mu$m luminosity 
for those LIRGs with an AGN detection. The galaxies are
color coded according to the AGN contribution to the integrated 
24\micron \, emission as in Fig.~3. The solid lines are the
relations found for Seyfert 1s (top, and $+ 1\sigma$ dispersion dashed
line) and Seyfert 2s (bottom and $-1\sigma$ dispersion) in the RSA sample 
\citep{Rigby2009}. The error bar on the upper left corner 
reflects the typical uncertainties in
the derived AGN bolometric luminosities from the comparison  in
Section~5.2,  and the typical uncertainty in the [O\,{\sc iv}] fluxes (see 
Section~3.2). } 
\vspace{0.5cm}
\label{fig_5}
\end{figure}

Because the [O\,{\sc iv}]$25.89\,\mu$m emission line  has
a  lower ionization potential (54.9\,eV) than the [Ne\,{\sc v}] lines
(97.1\,eV),  the mid-IR 
oxygen line is detected in a larger fraction of known AGNs 
\citep{Pereira2010lines}. Moreover, the \Oiv \, line appears to be a good
indicator of the AGN power \citep{Melendez2008, DiamondStanic2009, 
Rigby2009}, although it can
also be produced in star-forming galaxies \citep{Lutz1998, Pereira2010lines}. 

The \Oiv \, line is detected in 70\% of our sample of local LIRGs
(35/50 of the individual nuclei with IRS LH spectra available). In this
statistic we include NGC~3256 and IC~694 for which the line was
detected using IRS staring mode spectra by \cite{bernardsalas2009} and
\cite{AAH09Arp299}, respectively.
The detection rate in the GOALS\footnote{The GOALS sample contains a
  total of 248 individual LIRG nuclei in 
the {\it IRAS} RBGS, and thus our sample is part of it.}
LIRG sample \citep[53\% of their
sources, see ][]{Petric2011} is slightly lower than in our sample.
By comparison the \Oiv
\, line is detected in $\sim 26\%$ of the ULIRGs studied by
\cite{Veilleux2009}, but mostly in those galaxies 
classified as Seyfert and LINER. The detection of this line in the
\cite{Farrah2007} ULIRG sample is just under 50\%.  

In terms of the spectral class the \Oiv \, line
is detected in all (11) of  those galaxies optically classified as
AGN, that is, Seyferts and/or \Neva\, detections. Eleven of the 
17 galaxies optically classified as composites also show \Oiv
\, emission. 
Interestingly, eight LIRGs optically classified as H\,{\sc ii}-like
galaxies are detected in \Oiv, but have
no evidence of a hot dust component from the IRS AGN+SB decomposition
analysis.

\begin{figure*}
\center
\includegraphics[width=0.5\textwidth,angle=-90]{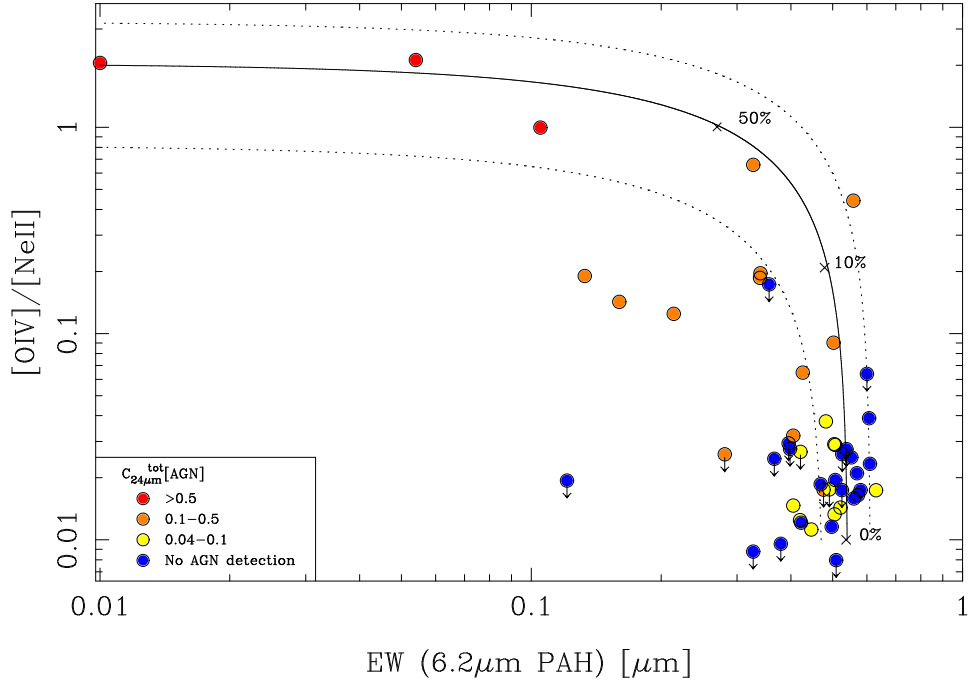}
\caption{\Oiv/\Neii \, line ratio versus EW(6.2\micron \, PAH) diagram
  for our 
sample of LIRGs, which are 
color coded according to the AGN contribution to the integrated 
24\micron \, emission, as in Figure~3. The solid line is a mixing
curve between pure 
star formation and pure AGN emission (see text for details), whereas
the dotted lines represent the $1\sigma$ dispersion in the  
\Oiv/\Neii \, line ratio for Seyfert galaxies and EW(6.2\micron
\, PAH) feature for purely star-forming galaxies. The percentage numbers on the
curves indicate the AGN contribution.}
\vspace{1cm}

\label{fig_7}
\end{figure*}

Figure~3  (upper panel) shows the distribution of the 
\Oiv \,  luminosities of our sample. We also show in
this figure as
filled histograms those 
LIRGs with an AGN detection, including Seyferts, composites,
 \Neva \, detections
and LIRGs with an AGN component from the AGN+SB decomposition. 
It is clear from this figure that all 
 the LIRGs with $L_{\rm [OIV]}\ge 10^7\,{\rm L_\odot}$ have an AGN
detected. For the merger LIRG NGC~3256 we plotted here our upper limit
from the {\it IRS} spectral mapping data, but the flux measurement of
\cite{bernardsalas2009} from an IRS staring mode 
spectrum provides a luminosity of $L_{\rm [OIV]}\sim 6 \times 
10^6\,{\rm L_\odot}$. At luminosities $\sim 10^5\,{\rm L}_\odot <
L_{\rm [OIV]}\le 10^7\,{\rm
L_\odot}$ approximately 50\% of the LIRGs are found to host an AGN.

As mentioned above the \Oiv \, line appears to be a good indicator of
the AGN power. For those LIRGs with an AGN  detection we
compare in Figure~4 $L_{\rm bol}({\rm AGN})$ from the IRS spectral
decomposition and the observed $L_{\rm [OIV]}$.  
The values for LIRGs with AGN detections are 
between those observed in Seyfert
1 and Seyfert 2 galaxies from the Revised Shapley-Ames (RSA) sample 
\citep{Rigby2009}, and thus consistent with the \Oiv \, emission being
produced mostly by the AGN. 

The \Oiv \, line can also be produced by star formation if the AGN
contribution to the total luminosity of the galaxy is below 5\%
\citep{Pereira2010lines}. This is indeed the case for non-AGN 
detections in our sample as we showed in Section~5.2. 
In Figure~3 (lower panel) we  compare the
observed \Oiv \, luminosities with those expected from pure
star-formation activity following 
\cite{Pereira2010lines}. We used the typical \Oiv/\Neii \, line
ratios for high metallicity starbursts \citep{Verma2003} 
and the empirical relation between the \Neii \, luminosity 
and  $L_{\rm IR}$ \citep{HoKeto2007} to predict the \Oiv \,
luminosity due to star formation. As can be seen from Figure~3, 
 all the galaxies with no AGN component detected
through the AGN+SB decomposition are consistent with pure star
formation. On the other hand, all the galaxies but one with $C_{24\mu{\rm
    m}}^{\rm tot}{\rm [AGN]} >0.1$ are more than $1\sigma$
above the relation expected for star formation. Moreover, for all galaxies
with $L_{\rm [OIV]}\ge 10^7\,{\rm L_\odot}$, this emission line is
mostly produced by the AGN, rather than by star formation.

\begin{figure*}
\center
\includegraphics[width=0.5\textwidth,angle=-90]{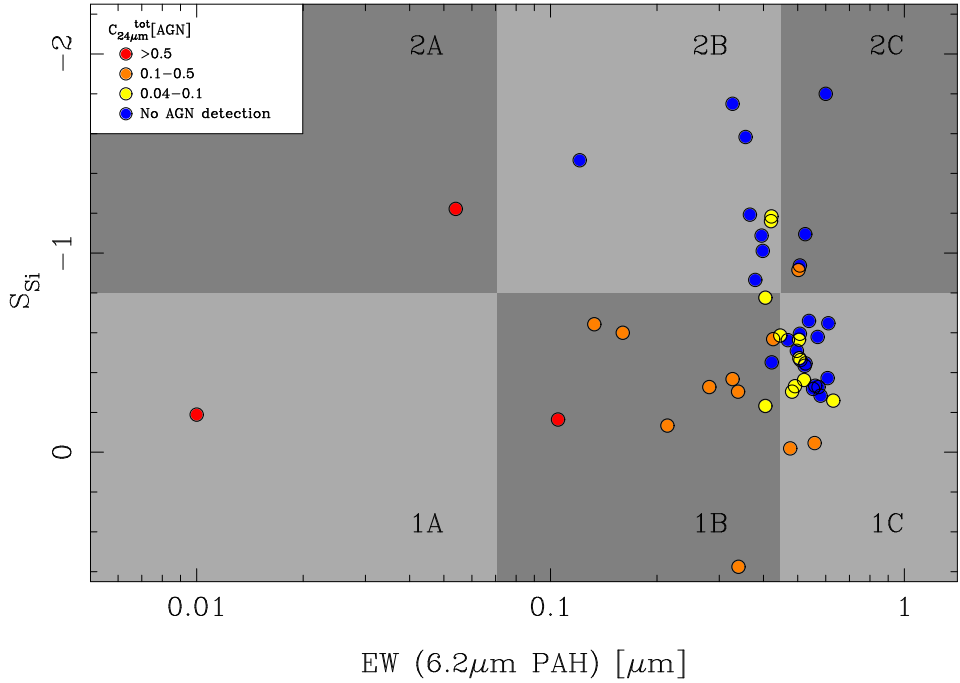}
\caption{Apparent strength of the silicate feature $S_{\rm Si}$
  versus EW(6.2\micron \, PAH) 
for our LIRGs color coded as in Figure~3. If $S_{\rm Si}$ is positive the 
  silicate feature is in emission, and if $S_{\rm Si}$ is negative the
  feature is in absorption. The regions shown in the diagram
  correspond to those defined by  \cite{Spoon2007} for their sample of
  ULIRGs. Class 1A is occupied by unobscured AGNs, class 1C by
  starburst galaxies (i.e., PAH-dominated galaxies), 
and class 1B by composite (AGN/SB activity)
  galaxies. Classes 2A through 2C are moderately obscured galaxies
  with increasing EW(6.2\micron \, PAH). Classes 3A through
  3C (not shown here) represent the most deeply embedded objects,
  although no ULIRGs are found in class 3C. } 
\vspace{0.8cm}
\label{fig_7}
\end{figure*}

\subsection{The [O\,{\sc iv}]$25.89\,\mu$m/[Ne\,{\sc ii]}$12.81$
  versus EW($6.2\,\mu$m PAH) Diagram} 

Diagnostic diagrams comparing the \Oiv/\Neii \, line ratio 
with the EW of one of the mid-IR PAH features  are commonly used to
assess the AGN   contribution to the IR emission of galaxies
\citep[see e.g.][]{Genzel98,  Dale2006}. In particular, this 
diagram has been used for 
  ULIRGs \citep{Farrah2007, Armus2007, 
Veilleux2009} and LIRGs \citep{Petric2011}.
This is because the \Neii \, line is expected to be produced mostly by
star formation, while the \Oiv \, line tends to be more luminous in
AGNs \citep{Pereira2010lines}. 
In this section we will assess if this diagram and the AGN
fraction of the total 24\micron \, emission provide consistent results.

Figure~5 presents this diagram for our sample, where the
galaxies have been color coded according to the AGN contribution at
24\micron. We also show in this diagram a mixing curve of pure star
formation and pure AGN emission. For the pure AGN emission we used the
median \Oiv/\Neii \, line ratio for Seyfert galaxies $2.0\pm 1.2$
\citep{Pereira2010lines} and EW(6.2\micron \, PAH)$=0.1\,\mu$m. For
the pure star formation we used the average value for high-metallicity
starbursts without AGN detections observed by {\it ISO} 
in the sample of \cite{Verma2003}, \Oiv/\Neii=0.011. We also use the
median value EW(6.2\micron \, PAH)$=0.54 \pm 0.07\,\mu$m (see Table~4)
of the HII-like LIRGs with no AGN detection 
for the pure star formation value.

It is clear from Figure~5 that the AGN fractions derived from this
diagram are entirely consistent with the AGN contribution to the total
24\micron \ emission. For instance, the three galaxies with $C_{24\mu{\rm
    m}}^{\rm tot}{\rm [AGN]} > 0.5$ have 
\Oiv/\Neii $>1$ and small values of the EW, which correspond to
$>50\%$ AGN fractions in the 
simplistic mixing curve plotted in this figure. Most LIRGs with $C_{24\mu{\rm
    m}}^{\rm tot}{\rm [AGN]} = 0.1-0.5$ have \Oiv/\Neii \, ratios 
 consistent with AGN fractions of $5-50\%$. Those LIRGs with
 small AGN contributions ($C_{24\mu{\rm
    m}}^{\rm tot}{\rm [AGN]} < 0.1$) tend to have values of the line ratio and
the EW consistent with those observed in nuclear and extra-nuclear
star-forming  regions \citep[e.g.][]{Dale2006}.
On the other hand, the observed value of EW(6.2\micron \, PAH) by itself 
is only found to be a good indicator of the nuclear
activity when the AGN contribution is dominant
$C_{24\mu{\rm
    m}}^{\rm tot}{\rm [AGN]} > 0.5$. For intermediate AGN
contributions there is a range of observed values of the EW of the
6.2\micron \, PAH feature, and thus deriving the AGN contribution is
not so readily done. 

\subsection{The Spoon et al. (2007) Diagram}
\cite{Spoon2007} put forward a diagram comparing the 
EW of the 6.2\micron \, PAH feature and the strength of the 9.7\micron \,
silicate feature to provide a general
classification of IR galaxies. This diagram has proven 
useful to assess the presence of an AGN not only for
local galaxies, but also for high-z IR bright galaxies
\citep[e.g.,][]{Farrah2008, HernanCaballero2009}. 
Figure~6 shows the Spoon et al. diagram for our sample of LIRGs, where
the galaxies are color-coded according to the AGN contribution at
24\micron. \cite{Pereira2010IRSmapping} presented this diagram
for the nuclei, integrated emission, and spatially resolved
measurements of those LIRGs in our sample observed in spectral mapping mode.

On the Spoon et al. diagram a large fraction (40\%) of the
LIRGs appear mostly concentrated in the region occupied by starburst
galaxies or class 1C.  Approximately 25\% are in class 1B, which is
where composite (AGN/SB activity) galaxies are located. Only one LIRG
is in the region of unobscured AGNs (Seyfert 1s and quasars) or class
1A, which typically tend to 
show low values of the EW of the PAH features
\citep{Roche91, Rigopoulou99, Buchanan2006, Tommasin2008}. The rest of
the sample, approximately 30\%, lie in those regions occupied by
moderately obscured nuclei ($-2 < S_{\rm Si}<-0.8$), classes  
2B and 2C. Only one LIRG is in region 2A. 
There are no deeply embedded ($S_{\rm Si}<-2$) 
nuclei among the LIRGs in our sample. 
On this diagnostic diagram ULIRGs are mostly located
along two branches. The first one is a horizontal branch that goes
from class 1A to class 1C, and the second is a
diagonal going from class 3A, which are deeply obscured nuclei (not
shown here), to class 1C \citep[see ][for more
details]{Spoon2007}. 

\begin{figure}
\center
\includegraphics[width=0.4\textwidth,angle=-90]{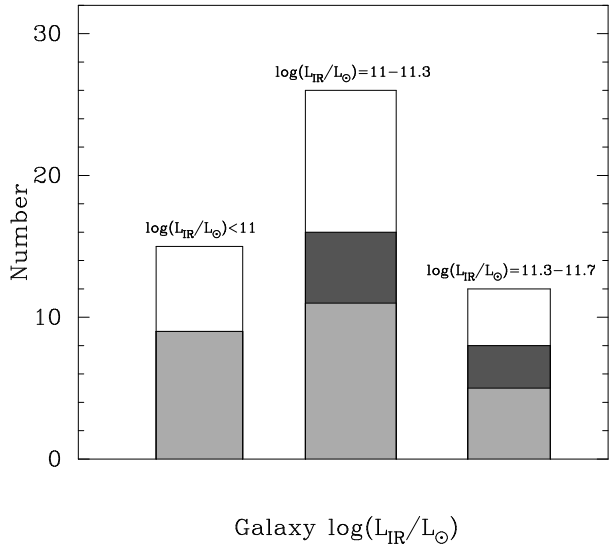}
\caption{Number of AGN detections (shaded histograms) as a function of the IR
  luminosity of the galaxy for the 53 individual nuclei of 
our volume-limited sample of local LIRGs (open histograms). 
The galaxies have been
divided in three luminosity bins. The light grey shaded
histograms correspond to detections based on the IRS spectral
decomposition, while the 
dark shaded histograms are LIRGs optically classified  as AGNs but not
detected from the AGN+SB decomposition.} 
\label{fig_7}
\end{figure}

The Spoon et
al. diagnostic diagram for LIRGs only provides a relatively good
separation between 
AGN-dominated LIRGs, LIRGs with intermediate AGN contributions and
LIRGs with very little or no AGN contribution. Only one of the
galaxies dominated by the AGN emission ($C_{24\mu{\rm
    m}}^{\rm tot}{\rm [AGN]} > 0.5$) appears in the 1A class,
whereas the other two are in the composite region and in the 2A
class. Most of those with intermediate AGN contributions ($C_{24\mu{\rm
    m}}^{\rm tot}{\rm [AGN]} = 0.1-0.5$) are in the 
composite region, as expected. However, two of them appear in region
1C, which is that occupied by 
PAH-dominated galaxies. Those with little 
($C_{24\mu{\rm
    m}}^{\rm tot}{\rm [AGN]} = 0.1-0.04$)
or no AGN contribution are
mostly either 
in the 1C region (starburst region), or near the border between class
2B and class 2C, and therefore are not easily identified as having an
AGN from this diagram.

\begin{figure*}
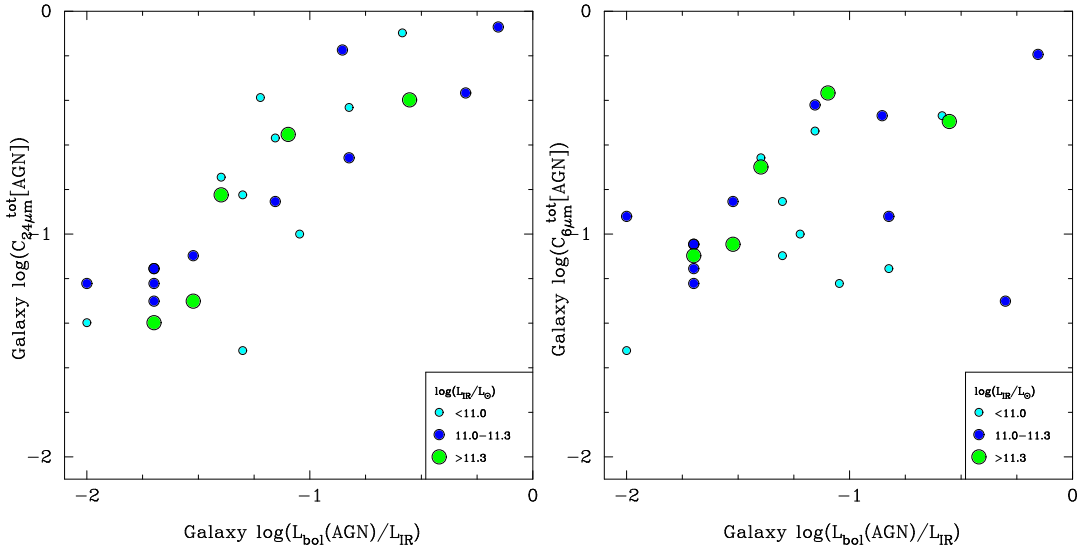

\center
\includegraphics[width=0.4\textwidth,angle=-90]{figure8a.ps}
\includegraphics[width=0.4\textwidth,angle=-90]{figure8b.ps}
\caption{{\it Left panel: }
Comparison between the AGN fraction at $24\,\mu$m and the
  ratio of the AGN bolometric luminosity to the IR luminosity of
  the galaxy for our sample of local LIRGs. 
%The AGN emission at
%  $24\,\mu$m is derived from the AGN+SB best fits as explained in 
%Section~3.1 (see also Figure~1), while the $24\,\mu$m total emission
%is measured from the {\it Spitzer}/MIPS $24\,\mu$m images. The AGN bolometric
%  luminosities are derived from the torus emission fitting since the
%  \cite{Nenkova2008a, Nenkova2008b} clumpy torus models scale directly
%  with the AGN   bolometric luminosity.   
The symbols are size (and color) coded according to the
  IR luminosity of the galaxy. {\it Right panel:} A similar comparison
  but at 6\micron. 
%In this case the total fluxes are measured from the
%{\it Spitzer}/IRAC 5.8\micron \, images.
}
\label{fig_9}

\vspace{0.85cm}
\end{figure*}

It is interesting to note here that the
\cite{Nenkova2008a, Nenkova2008b} 
  clumpy torus models never produce a very deep silicate
  feature (i.e., the models have $S_{\rm Si}\gtrsim -1$). This means 
that for deeply embedded galaxies the AGN templates used
  here \citep[see discussion by][]{Levenson2007}
may  not be appropriate and an AGN component might be
  missed. For instance two galaxies with relatively deep silicate
  features, IC~694 and MCG+02-20-003, are
  classified as LINER and composite, respectively, but an AGN
  component has not been detected through our spectral decomposition fit.
However, it is important to note here that deeply embedded sources are
rare among local LIRGs.

\section{Discussion}

\subsection{AGN Detection Rate in Local LIRGs}    
In this section we summarize the results of Sections~4 and 5 in terms of the
AGN detection rate in our complete volume-limited sample of local LIRGs and
investigate any possible dependence with the IR luminosity. We also compare the AGN detection rate in LIRGs with other types of galaxies.

Out of the
50 galaxies with IRS spectroscopy we detected an AGN component in 25
using the AGN+SB spectral decomposition  (see Table~2, and Figures~1 and A1). 
An additional  eight galaxies are classified spectroscopically as
composite or LINER. Of these, one did not have IRS spectroscopy (the IRS
spectrum of NGC~1614 was not well centered)
and for the remaining seven 
the AGN was not detected based on our spectral decomposition 
(UGC~1845, MCG~+02-20-003, IRAS~17138$-$1017, IC~694, NGC~5734, 
NGC~7591, Zw049.057). This brings the combined optical+IR AGN
detection rate in the individual nuclei of our sample of LIRGs to
$\sim 62\%$. In terms of the 45 {\it IRAS} systems, an AGN is found in 32,
that is, $\sim 70\%$. This AGN detection rate is in good agreement with that
derived from optical emission lines \citep{Yuan2010} and X-ray
emission \citep{Risaliti2000, Pereira2011a}.

Figure~7 shows the  number of AGN detections for the
individual nuclei of our sample of LIRGs in three IR luminosity bins. 
As can be seen from this figure, the AGN detection rate stays
approximately constant at a $\sim 60-65\%$ rate for the three luminosity
bins $\log (L_{\rm IR}/\,{\rm L}_\odot) < 11$, $\log (L_{\rm
  IR}/\,{\rm L}_\odot) = 11-11.3$ and $\log (L_{\rm IR}/\,{\rm
  L}_\odot) = 11.3-11.7$.

\cite{Petric2011} have conducted a study similar to ours using the
GOALS sample of 248 LIRGs. They identified AGN through \Neva \,
detections in 18\% of their sample, virtually identical to the
22\% detection rate for us. They also detected \Oiv \, in
53\% of their galaxies, whereas we detected it in 70\% (see
Section~5.3). They do not 
count the [O\,{\sc iv}] detections as AGN, whereas we find that many of these
galaxies do indeed harbor active nuclei (from the spectral
decomposition and/or optical class, see Section~5.3).
\cite{Petric2011} also used the EW of the 
$6.2\,\mu$m aromatic feature and the 5 to $15\,\mu$m continuum flux ratios
as AGN indicators, and concluded that 
$\sim 10\%$ of their sample are dominated by AGN in the
mid-IR. We find that $\sim 6\%$ (Table~4) 
of our sample have a mid-IR 
AGN contribution greater than 50\% of the total luminosity
at 24\micron. The difference may arise
from slightly different methodologies, but both studies agree that the
AGN plays a substantial role in the mid-IR energetics only in a
relatively small minority of these galaxies. However, we also find
that 50\% of the galaxies in our sample contain AGNs, based on our
mid-IR spectral decomposition method, and $\sim$ 60\% if we combine
optical and mid-IR indicators. Our ability to find many more AGNs than
\cite{Petric2011} did in the larger but similar sample suggests
that our spectral decomposition method is a powerful way to identify
subtle AGN features. As a result, we can show that
AGNs accompany star formation activity in a large proportion of local
LIRGs, although in most
cases the AGNs are not energetically important.

The AGN detection rate in local LIRGs is very similar to that of local ULIRGs.
The latter is found to be 70\% on average  
using mid-IR diagnostics \citep{Imanishi2007, Nardini2010}.
In lower luminosity galaxies AGNs are detected in
much smaller fractions. For instance, in the optically-selected RSA
sample \cite{Maiolino95} found a  $5-16\%$ AGN incidence, based on
optical indicators. The AGN fraction in nearby ($d<15\,$Mpc) 
moderately IR luminous
galaxies ($L_{\rm IR} \gtrsim 10^9\,{\rm L}_\odot$) is slightly
higher, 27\%, when
including \Neva \, detections, as demonstrated by \cite{Goulding2009}.

\subsection{AGN Bolometric Contribution in Local LIRGs}
Although the combined optical+IR 
AGN detection rate in local LIRGs is relatively high,
$\sim 62\%$, and only slightly less than in local ULIRGs, the
important quantity is the AGN bolometric contribution to the IR
luminosity of the galaxies. 
Using  the AGN bolometric luminosities derived from the spectral
decomposition (Section~5.2), we can derive the AGN bolometric contribution 
to the total IR luminosity of local LIRGs (both with and without AGN
detections). We find that AGNs 
contribute $5^{+8}_{-3}$\%  of the IR luminosity\footnote{Strictly
  speaking, this is just an upper 
limit to the AGN contribution 
because we are assuming that in LIRGs $L_{\rm bol}({\rm AGN}) \simeq
 L_{\rm IR}({\rm AGN})$.}. The upper and lower limits take into account the
uncertainties in the derived AGN bolometric luminosities. For the 9
galaxies with an AGN optical classification 
but no AGN emission detected from the
AGN+SB decomposition, we assumed the typical  
AGN bolometric luminosity of the composite sources (see
Section~5.2). This estimate proves that the bulk of
the IR luminosity of local LIRGs is due to star formation activity. 

\begin{figure*}
\center
\includegraphics[width=0.4\textwidth,angle=-90]{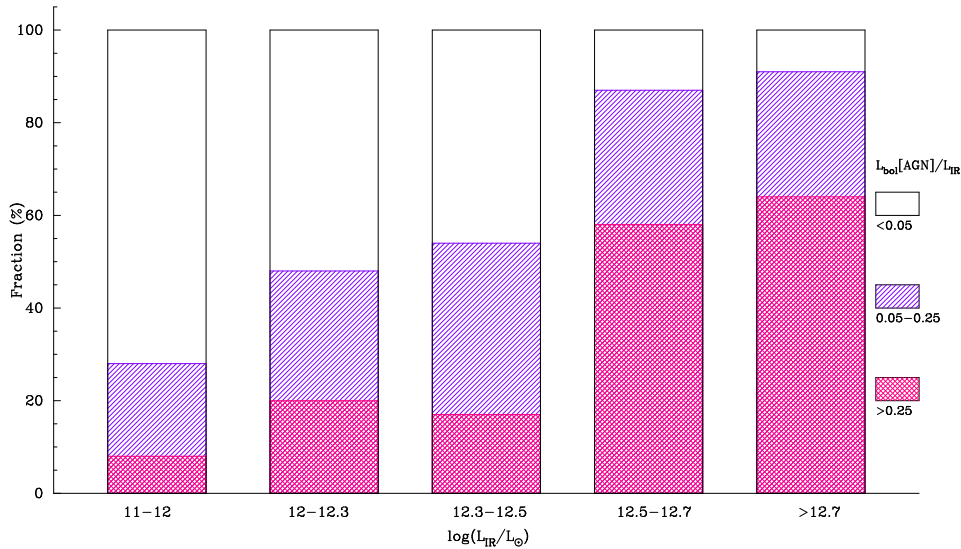}
\caption{Summary of the fractional AGN bolometric contribution to the IR
  luminosity for the local
  sample of LIRGs and comparison with local ULIRGs from \cite{Nardini2010}.} 
\label{fig_10}
\vspace{0.75cm}
\end{figure*}

Our estimate of the AGN contribution to the total IR
luminosity of LIRGs is in
excellent agreement with that estimated from the X-ray properties of
the representative sample studied by \cite{Pereira2011a}. 
\cite{Petric2011}
estimated that  AGN supply $\sim 12\%$ of the total energy emitted by
the GOALS LIRGs, which is compatible with our upper limit of the AGN contribution.  

Taking the newly determined IR luminosity density produced by LIRGs in
the local universe of  $\Omega_{\rm IR } = 6\times10^6 L_\odot \,{\rm
  Mpc}^{-3}$ \citep{Goto2011}, we estimate that the AGN IR 
luminosity density in the LIRG luminosity range is  
 $\Omega_{\rm IR }^{\rm AGN} = 3_{-2}^{+5}\times10^5 L_\odot \,{\rm
  Mpc}^{-3}$. Our estimate of the AGN IR luminosity density is
about seven times lower than that of \cite{Goto2011}, probably because
these authors assumed that in those LIRGs classified as active
galaxies and especially those classified as composite the IR
luminosity is entirely due to the AGN.

Finally, we can also compare $C_{6\mu{\rm
    m}}^{\rm tot}{\rm [AGN]}$ and $C_{24\mu{\rm
    m}}^{\rm tot}{\rm [AGN]}$ with the AGN bolometric contribution to
the IR luminosity of  the galaxies (Figure~8).
The relation between the AGN contribution in the mid-IR and
the AGN bolometric contribution is slightly better at 24\micron \,
than at 6\micron. As explained 
above, the IRAC 5.8\micron \, images of some LIRGs may include a
sizable contribution from the 6.2\micron \, PAH feature, and/or the
6\micron \, emission of type 2 AGNs might not  be
fully isotropic. It can also be seen from Figure~8 that, although $C_{24\mu{\rm
    m}}^{\rm tot}{\rm [AGN]}$ and $L_{\rm bol}[{\rm AGN}]/L_{\rm IR}$
do not follow a 1:1 relation,  the relation is nearly linear. 
On average $C_{24\mu{\rm
    m}}^{\rm tot}{\rm [AGN]} = (2.9 \pm 1.6) \times L_{\rm bol}[{\rm
  AGN}]/L_{\rm IR}$. We conclude that estimating the AGN fraction at
24\micron \, is a good proxy for the bolometric AGN contribution in
LIRGs. Moreover, because the fractional AGN contribution 
is on average higher in the mid-IR than bolometrically, 
the mid-IR spectral range is very appropriate to 
study the fractional AGN/SB contribution. \cite{Wu2011} reached a
similar conclusion for a sample of IR luminous galaxies at
$z\sim 0.3$.

\subsection{Comparison with Local ULIRGs and high-$z$ IR bright galaxies}

We summarize the AGN bolometric contribution for our sample of local
LIRGs (with and without AGN detections) in Figure~9. We also compare
in this figure our estimates for local LIRGs with those for local
ULIRGs of \cite{Nardini2010}. This comparison is meaningful because
both our work and that of Nardini et al. quantify the AGN
bolometric luminosities. In this figure we separated the 
AGN bolometric contributions $L_{\rm bol}[{\rm
  AGN}]/L_{\rm IR}$ in three ranges:  $<0.05, \,
0.05-0.25, \,>0.25$.  
In approximately one-third of local LIRGs the AGN bolometric
contribution is mild, $L_{\rm bol}[{\rm AGN}]/L_{\rm IR}\ge
0.05$. Only $\simeq 8\%$ of local LIRGs have
a significant  AGN 
contribution, $L_{\rm bol}[{\rm AGN}]/L_{\rm IR}>0.25$.

It is clear from Figure~9 that  the
fraction of galaxies with $L_{\rm bol}[{\rm AGN}]/L_{\rm IR}>0.05$
increases at higher $L_{\rm IR}$, going from 30\% in local LIRGs to
$\sim 50\%$ to ULIRGs  with $L_{\rm 
  IR}<3\times 10^{12}\,{\rm L}_\odot$. However, it is only at $L_{\rm
  IR}>5\times 10^{12}\,{\rm L}_\odot$ that the AGNs 
start dominating bolometrically the IR luminosity in a large fraction
of local ULIRGs; 40\%  of ULIRGs in the high luminosity bin in
Figure~9 have  $L_{\rm bol}[{\rm AGN}]/L_{\rm IR}>0.60$ 
\citep[see Figure~9 of ][]{Nardini2010}.

In summary, in the local universe 
there is an increasing bolometric contribution from
AGNs  at higher IR luminosities, going from 5\% in LIRGs to 
an average of 27\% in ULIRGs \citep{Nardini2010} 
and $35-40\%$ for the 1\,Jy ULIRG sample \citep{Veilleux2009}. 

While the trend for an increased AGN bolometric dominance at high IR
luminosities is well established locally, at high-$z$ 
it is still a matter of debate. For instance,
\cite{Valiante2009} needed to introduce an evolution in the AGN 
contribution to model the submillimeter to mid-IR number counts 
and redshift distribution of high-$z$ IR galaxies. In particular they
found that at a given IR luminosity high-$z$ IR galaxies 
needed a smaller AGN contribution than locally. This is in line with results of
\cite{Fadda2010} using deep mid-IR spectroscopy of 
$z\sim 1$ LIRGs and $z\sim 2$ ULIRGs.  In both populations the fraction of AGN
dominated sources is small, being  $\sim 12\%$ in ULIRGs and $\sim
5\%$ in LIRGs. Also \cite{Wu2011} found no significant
change in the overall star formation contribution to $L_{\rm IR}$ from LIRGs
to ULIRGs at $z\sim 0.3$. However, other works found evidence for higher AGN
contributions for the most IR luminous galaxies at high-$z$. Indeed, 
\cite{MenendezDelmestre2009} found 
that, although submillimeter galaxies are dominated by intense star
formation, the average AGN bolometric contribution could be as high
30\%. Moreover, \cite{Rujopakarn2011} found a strict
upper limit of $\sim 5\times 10^{12}-10^{13}\,{\rm L}_\odot$ 
for the IR luminosity  to be due to star formation in galaxies 
at all redshifts. Above that limit, AGN are expected to provide the power.

\section{Conclusions}\label{s:conclusions}
We have decomposed the {\it Spitzer}/IRS $\sim 5-38$\micron \, 
spectra of a complete
volume-limited ($d\sim 40-78\,$Mpc) 
sample of local LIRGs into AGN and SB components.
The main goal of this work is to quantify the AGN 
contribution to the mid-IR and 
 IR emission of these systems. For the SB component we used starburst
and LIRG templates, whereas the AGN component was fitted with the
\cite{Nenkova2008a, Nenkova2008b}
clumpy torus models. We have also compared our mid-IR AGN
detections with this method with other AGN indicators, including
mid-IR spectral features, optical classification and X-ray
properties. Our main conclusions can be summarized as follows, 
\begin{enumerate}
\item Using the AGN+SB spectral decomposition we detected an AGN
  component in 25 
  out of the 50 individual LIRG  nuclei
with IRS spectroscopy. For an additional nine galaxies 
optically classified as composite (AGN/SB) we did not 
detect an AGN  with our
method. The combined optical and mid-IR  AGN
detection rate is $\sim 62\%$ (33/53) 
for the individual nuclei of local LIRGs. This is in good agreement with
the AGN detection rate obtained from optical spectroscopy, if composite
objects do contain an AGN.
\item The AGN contribution to the mid-IR continuum emission within the
  IRS slits is small, and decreases toward longer wavelengths. The 
median values range from $C_{6\mu{\rm m}}^{\rm IRS}{\rm [AGN]}=0.30$,
and $C_{24\mu{\rm m}}^{\rm IRS}{\rm [AGN]}=0.18$  to $C_{30\mu{\rm
    m}}^{\rm IRS}{\rm [AGN]}=0.11$. 
\item We used IRAC 5.8\micron \, and MIPS 24\micron \, images to
  derive the AGN contribution to the total emission at these 
mid-infrared wavelengths. The
  median AGN contributions for those LIRGs with an AGN detection are
 $C_{6\mu{\rm m}}^{\rm tot}{\rm [AGN]}=0.12$ and 
$C_{24\mu{\rm m}}^{\rm tot}{\rm [AGN]}=0.15$,  with only $\sim 6\%$
(3/50) of local LIRGs having a dominant AGN contribution at 
24\micron \, ($C_{24\mu{\rm m}}^{\rm tot}{\rm [AGN]} > 0.5$).
\item We detected the \Oiv \, high excitation 
emission line in 70\% (35/50) of
  the individual LIRG nuclei. All the galaxies in our sample 
with  $L_{\rm [OIV]}\ge 10^7\,{\rm L_\odot}$ contain an AGN and
 these luminosities are consistent with those expected from their
 derived AGN bolometric luminosities. On the other hand, the 
\Oiv \, luminosities of those
 galaxies without an AGN detection can be explained as produced by 
star formation activity.
\item When compared with other mid-IR spectral AGN diagnostics, we
  found that our 24\micron \, AGN fractional components are consistent
  with those   derived from the  
 [OIV]/[NeII] \, versus EW(6.2\micron \, PAH) diagram. Using the Spoon et
 al. diagram we can identify the presence of an AGN when 
$C_{24\mu{\rm m}}^{\rm tot}{\rm [AGN]} \gtrsim  0.1$.
\item From the scaling of the fitted torus models to the AGN component
  of the IRS spectra we derived  AGN bolometric luminosities in the
  range $L_{\rm bol}{\rm (AGN)}=0.4 -50\times 10^{43}\,{\rm erg \,
    s}^{-1}$ with a median value of $L_{\rm bol}{\rm (AGN)}=2\times
  10^{43}\,{\rm erg \,  
    s}^{-1}$. These are in a fairly good agreement with those
  estimated from hard X-ray measurements after applying a bolometric
  correction. 
\item One-third of local LIRGs
have $L_{\rm bol}[{\rm AGN}]/L_{\rm IR}\ge 0.05$, with only $\simeq 8\%$ having
a significant  AGN 
contribution $L_{\rm bol}[{\rm AGN}]/L_{\rm IR}>0.25$. This is in contrast 
with the $\sim 20\%$  of local ULIRGs  with $L_{\rm
  IR}<3\times 10^{12}\,{\rm L}_\odot$ and the $\sim 60\%$ of ULIRGs 
with $L_{\rm
  IR}> 3\times 10^{12}\,{\rm L}_\odot$ having $L_{\rm bol}[{\rm
  AGN}]/L_{\rm IR}>0.25$, respectively. 
\item Adding up the AGN bolometric luminosities in our sample
 we find that  AGNs are only responsible for 
$5^{+8}_{-3}$\% of the total IR luminosity produced by local LIRGs.
This translates into an
AGN IR luminosity density of 
$\Omega_{\rm IR }^{\rm AGN} = 3_{-2}^{+5}\times10^5 L_\odot \,{\rm
  Mpc}^{-3}$ in local LIRGs. Our results prove that the bulk of
the IR luminosity of local LIRGs is due to star formation activity. 

\end{enumerate}

In summary, mid-IR spectral decomposition is a powerful tool to
estimate the AGN contribution to both the mid-IR emission and the
total emission of IR-selected sources. This is because not only are we
able to identify all Seyfert-like AGN in local LIRGs but also because 
this technique is
powerful enough to identify subtle AGN features (i.e., low luminosity
AGN) that might be missed by other mid-IR methods. As a result, 
we have shown that 
AGNs accompany star formation activity in the large majority of local
LIRGs, although in most
cases the AGNs are not energetically important.

\section*{Acknowledgements}

We thank Emanuele Nardini, Guido Risaliti, and Jane Rigby
 for sharing some of their
results with us. We are very grateful to Moshe Elitzur for the {\it
  CLUMPY} torus models and to Andr\'es Asensio Ramos and Cristina 
Ramos Almeida for
their continuing work in developing the BayesClumpy fitting tool. We
thank Cristina Ramos Almeida and 
Pilar Esquej for a careful reading of the manuscript and for their
suggestions. A.A.-H. is grateful to the Astrophysics Department, 
Oxford University, where part of this research was conducted, 
for their hospitality. 
We thank the referee for useful suggestions
and comments.

A.A.H and M.P.-S. acknowledge support from the Spanish Plan Nacional
de Astronom\'{\i}a y Astrof\'{\i}sica under grants AYA2009-05705-E 
and AYA2010-21161-C02-1. 
M.P.-S. acknowledges support from the CSIC under grant JAE-Predoc-2007. 

This research has made use of the NASA/IPAC Extragalactic Database
(NED) which is operated by the Jet Propulsion Laboratory, California
Institute of Technology, under contract with the National Aeronautics
and Space Administration.

\appendix
\subsection*{A1. The CLUMPY Torus Models}
We used the \textit{CLUMPY} models \citep{Nenkova2008a, Nenkova2008b} 
to represent the AGN continuum emission in the mid-IR. These models 
are described by  six parameters
that deal with the torus geometry and cloud properties. These are: (1)  
the radial thickness of the torus  $Y = R_{\rm o}/R_{\rm d}$, where
$R_{\rm o}$ and $R_{\rm d}$ are the outer and inner radii of the
torus, respectively. The inner
radius of the torus in these models 
is set by the dust sublimation temperature, which is 
assumed to be $T_{\rm sub} \approx 1500$ K. (2)  
The angular distribution of the clouds, which is assumed to have a smooth
boundary, is described as a Gaussian with a width
parameter $\sigma_{\rm torus}$. (3) The radial distribution of the
clouds, which is described as a declining
power law with index $q$ ($\propto r^{-q}$). 
(4) The mean number of clouds along  a radial equatorial ray 
$N_0$. (5) The optical depth of the clouds
$\tau_V$. (6) The viewing angle to the torus, $i$. 
In the models the radiative transfer equations are solved for each
clump and thus the solutions depend mainly on
the location of each clump within the torus, its optical depth, and
the chosen dust composition. For the fits done in this work we
 adopted a dust extinction profile
corresponding to a standard cold oxygen-rich ISM  dust \citep{Ossenkopf92}. The total torus emission is
calculated by integrating the source function of the total number of
clumps convolved with the radiation propagation probability along the
torus \citep{Nenkova2002}. For unobscured views of the AGN, it is also
possible to include  its contribution to the resulting IR
emission. The AGN continuum emission in 
these models is characterized with a piecewise power law distribution
\cite[see][for details]{Nenkova2008a}. 
In addition to the the six torus  model parameters, there is an 
extra parameter to account for the vertical displacement
needed to match the fluxes of a given model to the observations. 
This vertical shift, which we allow to vary freely,  scales with
the AGN bolometric luminosity \citep[see][]{Nenkova2008b}.

Given the large number of 
\textit{CLUMPY} models (currently more than $10^6$) and the inherent degeneracy
in the torus model parameters, \cite{BC} took a Bayesian approach to
fit models to the data and to derive meaningful confidence levels for
the fitted parameters. To this end, they developed a tool called
BayesClumpy that allows fitting both photometric points and/or mid-IR
spectra. For the Bayesian inference of 
BayesClumpy  we used an interpolated version  of 
the {\it CLUMPY} dusty torus models. 
We refer the reader to \cite{BC}  for details on the
interpolation methods and algorithms 
used by BayesClumpy to perform the torus model fits to the data.

\subsection*{A2. AGN+SB Spectral Decomposition Fits}
We used an iterative method to perform the AGN+SB decomposition. In
the first step we started by using only two AGN templates given the large
number of \textit{CLUMPY} torus models and their degeneracy 
\citep[see][for detailed discussions on this issue]{RA09, RA11,
  BC}. We used the best  fit \textit{CLUMPY} torus model of the
average spectral energy distributions of Seyfert 1s and Seyfert 2s
inferred by \cite{RA11}.  
We obtained the best fit to the IRS spectra for each LIRG in our sample
with this initial combination of AGN+SB templates. We next subtracted
the SB component from the original IRS spectrum to obtain
 the AGN-only mid-IR spectrum. This AGN-only spectrum was then
fitted with the BayesClumpy routine \citep{BC} to infer the best fit torus model
and the corresponding scaling for each galaxy. We finally refitted the observed
IRS spectrum of the LIRG using its corresponding best-fit torus model, allowing for rescaling, 
and the appropriate SB template that minimized $\chi^2$. The fits were
done in the $6-30\,\mu$m spectral range to avoid the edges of the
spectra and the slightly decreased signal-to-noise ratio of some of
the spectra observed in mapping mode at $\lambda > 30\micron$. We
note, however, that the fits tend to be good also at $\lambda >
30\micron$. We considered an AGN detection when its contribution was
greater than 5-7\% at 20\micron. Below this limit we were not able to
fit the AGN-only spectra with BayesClumpy. This is because
after subtracting the SB template in these cases a large proportion
of the data points were close to zero or even negative. 
For each fit the formal errors of the AGN fractional contribution within
the IRS slit were calculated using the observational errors of the 
IRS spectra and the covariance matrix. The typical errors of
the AGN fractional contribution within the IRS slit at $6\,\mu$m are 
$0.01-0.03$. If the models fit the data, as is the case for most
galaxies, these can be taken as the typical uncertainties of
the AGN fractional contributions. This is because the variation of the fitted 
parameters is compatible with the data observational errors. The
uncertainties
associated with the use of different SB templates are discussed briefly in Section~4.1.

We present the best AGN+SB fits for all the
LIRGs in our sample in Figures~1, A1, and A2. As can be
seen from Figure~A2 a few LIRGs deemed not to have a mid-IR AGN
detection have best fit requiring a small AGN contribution. These are NGC~5734,
which is optically classified as composite, and ESO~221$-$G010 and
IC~4734, which
are classified as HII (see Table~1).
The AGN fraction at $20\micron$ for all of them was, however, below the imposed 
limit. In these 
cases we were not able to fit the AGN-only
emission with the torus models, and the AGN template corresponds to
the Seyfert 2 template used in the first step of the iteration.

\begin{figure*}
\includegraphics[width=0.32\textwidth]{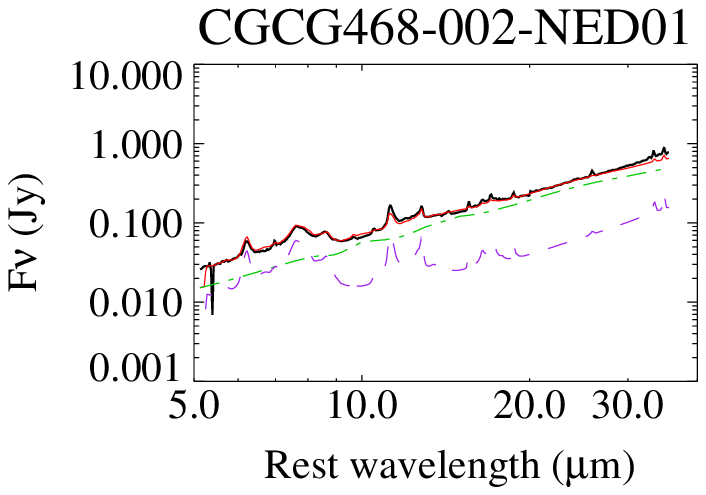}
\includegraphics[width=0.32\textwidth]{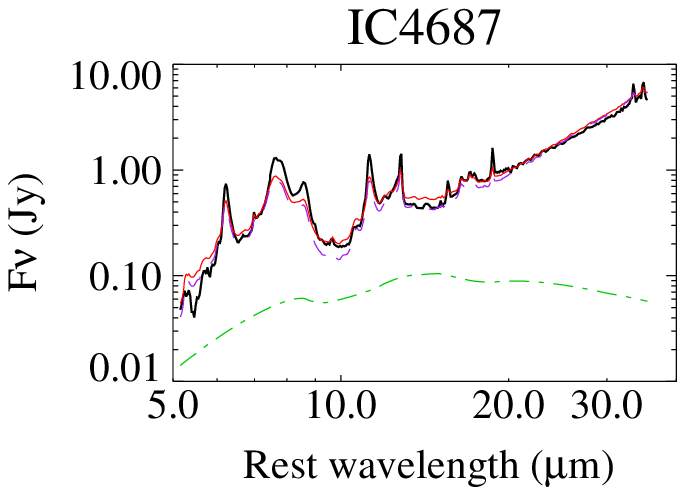}
\includegraphics[width=0.32\textwidth]{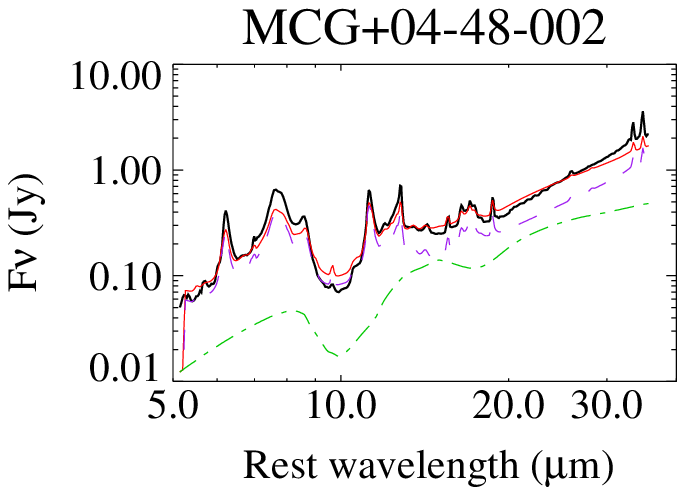}

\includegraphics[width=0.32\textwidth]{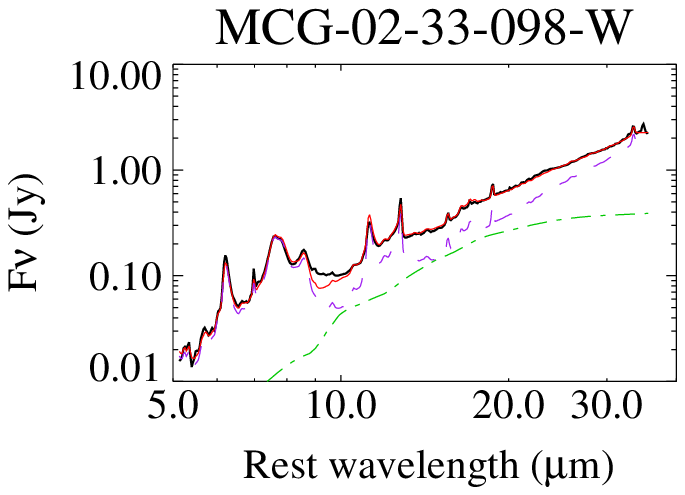}
\includegraphics[width=0.32\textwidth]{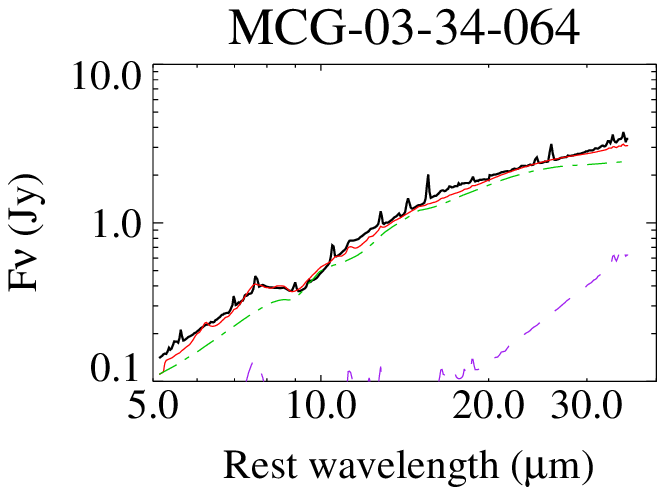}
\includegraphics[width=0.32\textwidth]{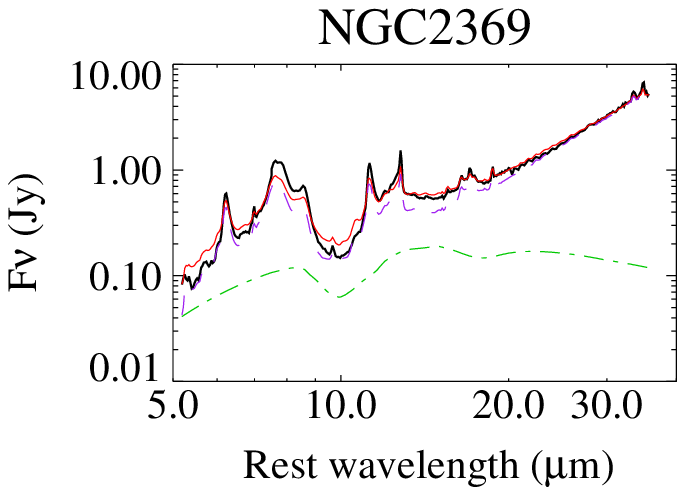}

\includegraphics[width=0.32\textwidth]{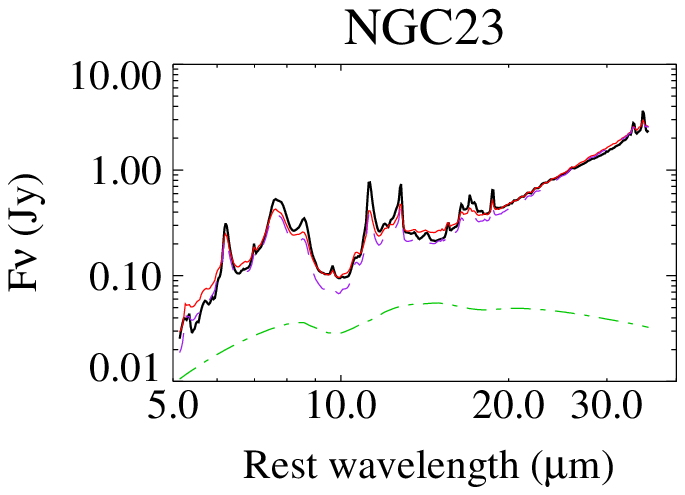}
\includegraphics[width=0.32\textwidth]{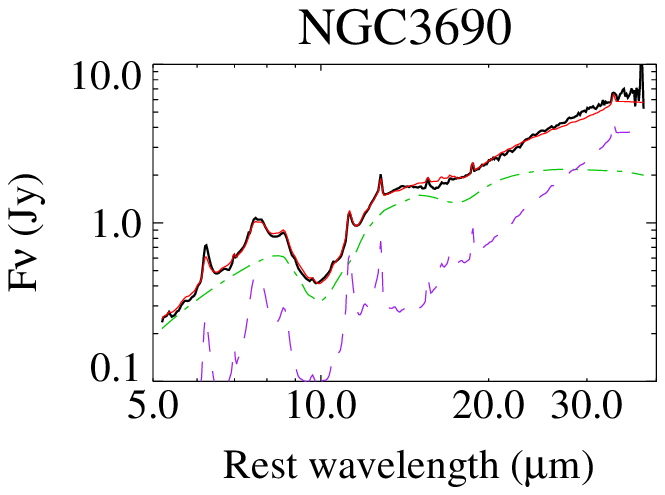}
\includegraphics[width=0.32\textwidth]{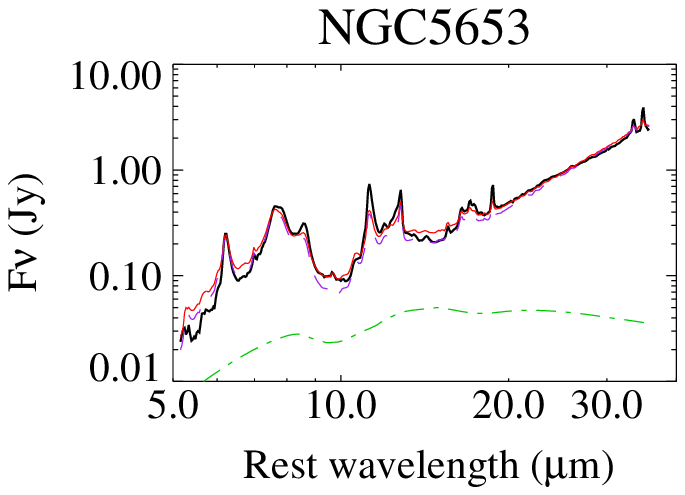}

\includegraphics[width=0.32\textwidth]{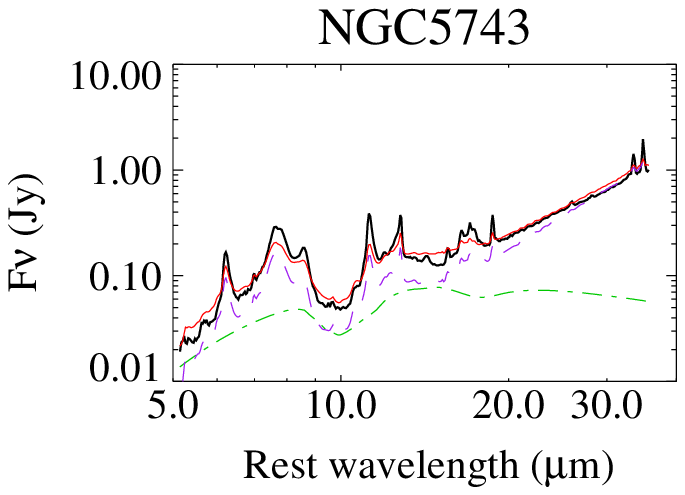}
\includegraphics[width=0.32\textwidth]{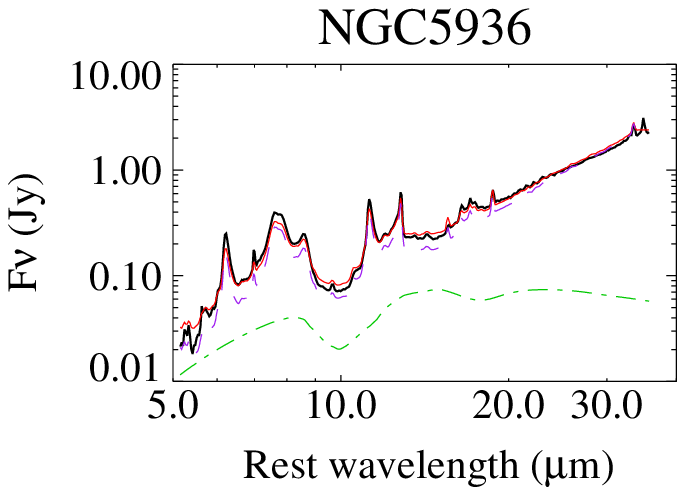}
\includegraphics[width=0.32\textwidth]{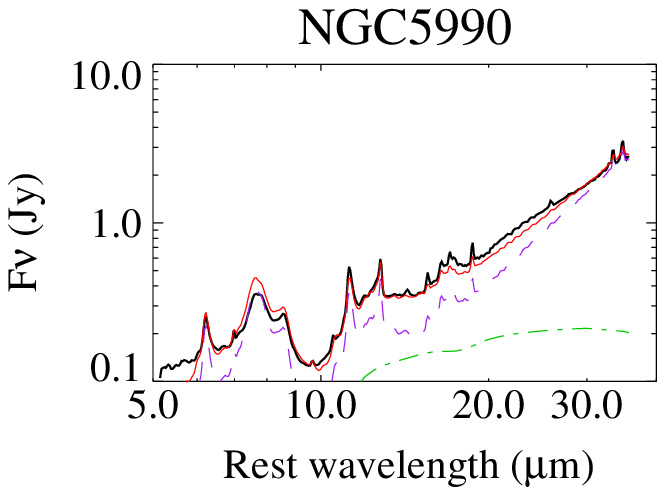}

\includegraphics[width=0.32\textwidth]{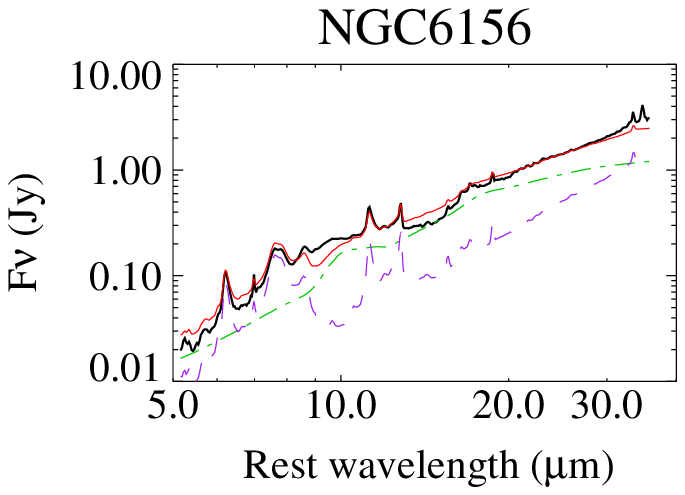}
\includegraphics[width=0.32\textwidth]{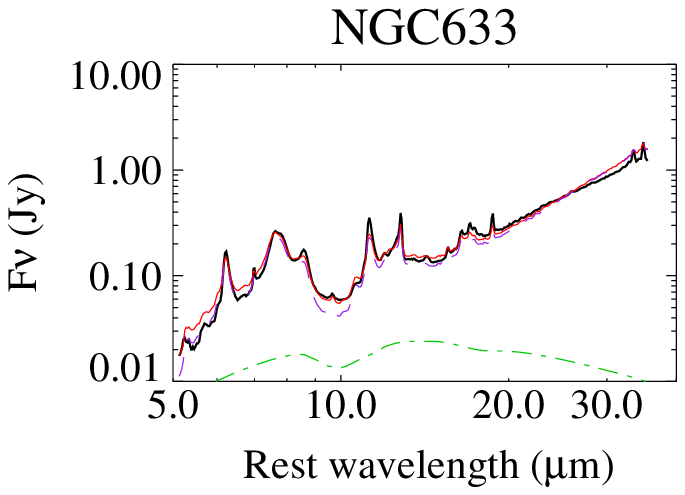}
\includegraphics[width=0.32\textwidth]{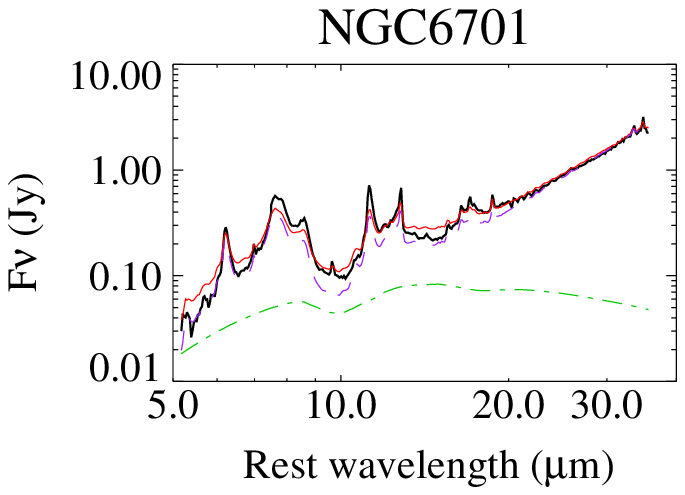}

\includegraphics[width=0.32\textwidth]{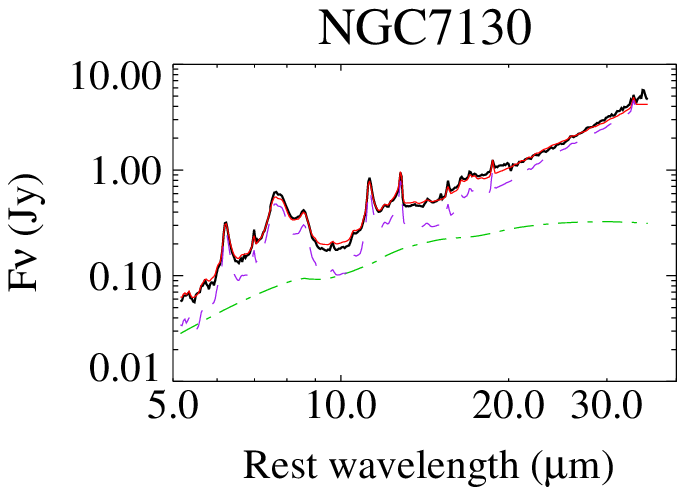}
\includegraphics[width=0.32\textwidth]{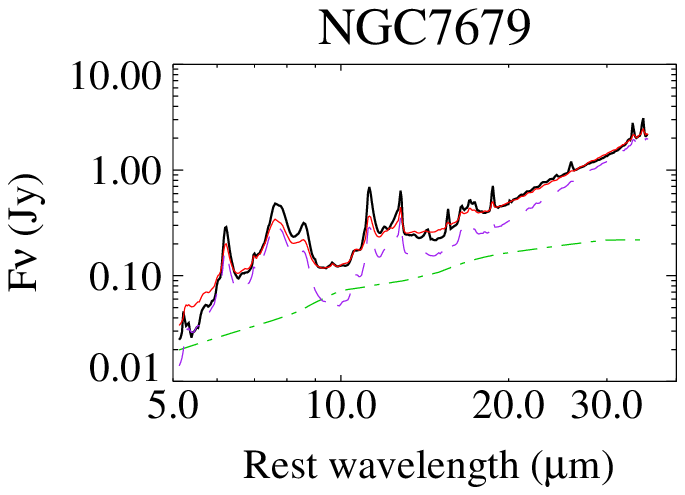}
\includegraphics[width=0.32\textwidth]{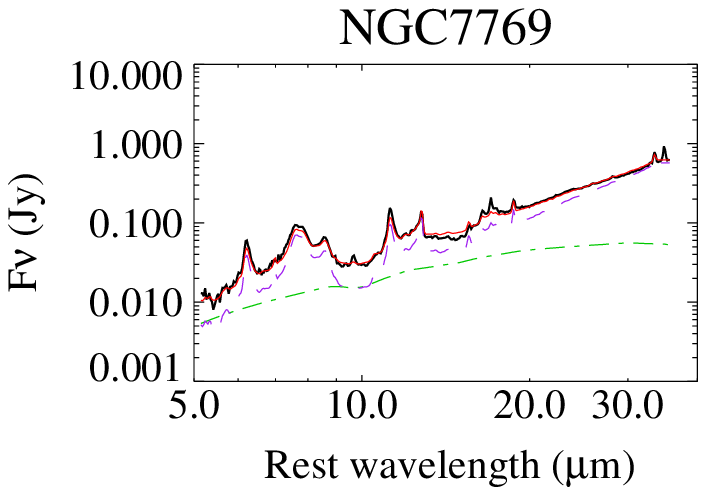}

 {{\bf Figure~A1}. AGN+SB decomposition fits for those LIRGs with a mid-IR AGN
  detection. Lines are as in Figure~1.}
\end{figure*}

\begin{figure*}
\includegraphics[width=0.32\textwidth]{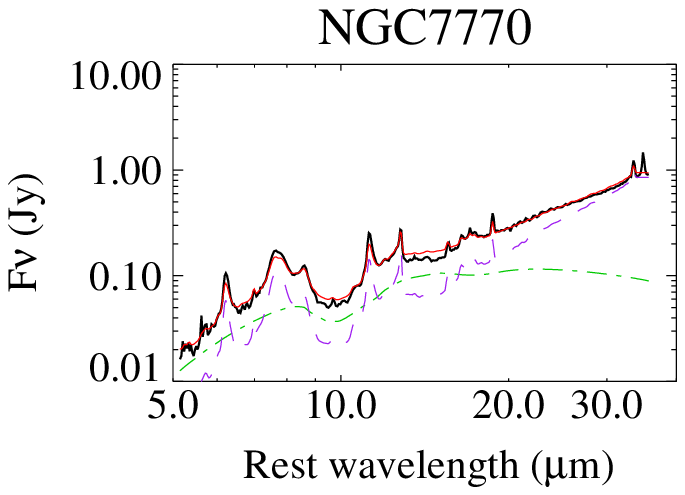}
\includegraphics[width=0.32\textwidth]{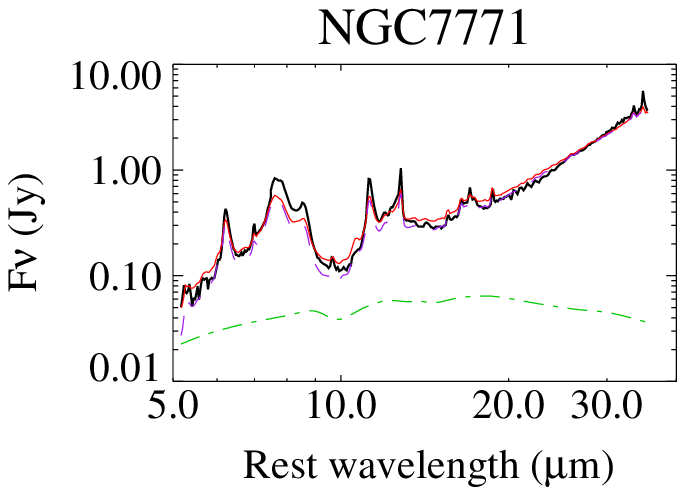}
\includegraphics[width=0.32\textwidth]{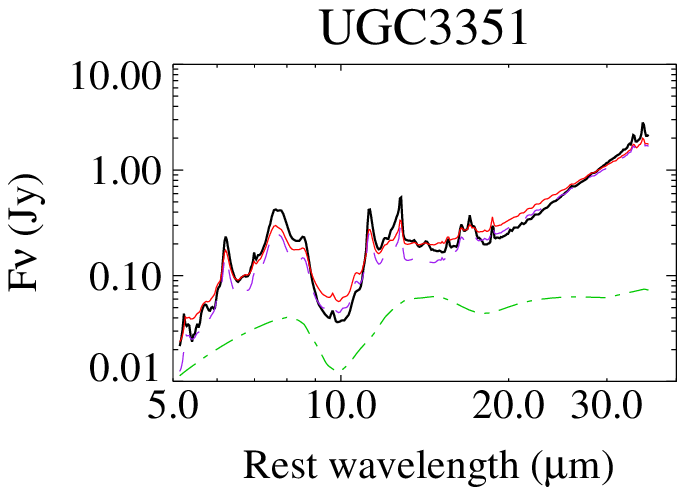}

\includegraphics[width=0.32\textwidth]{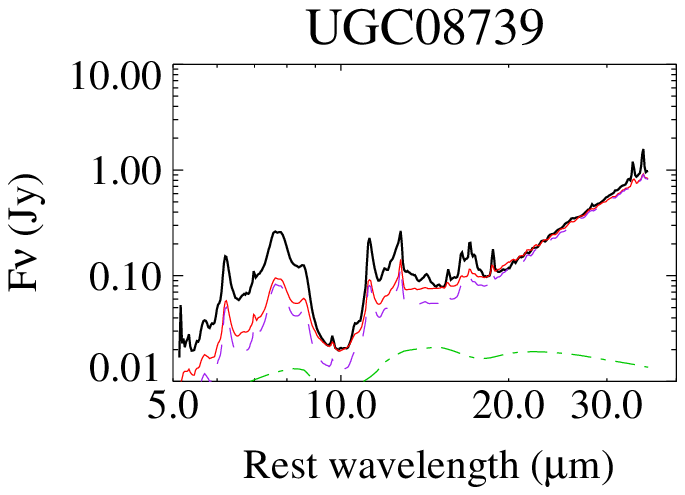}

{{\bf Figure~A1}. Continued.}
\end{figure*}

\begin{figure*}
\includegraphics[width=0.32\textwidth]{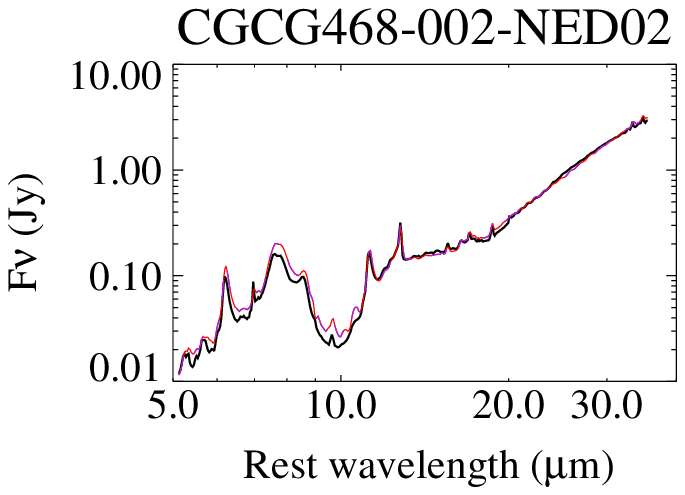}
\includegraphics[width=0.32\textwidth]{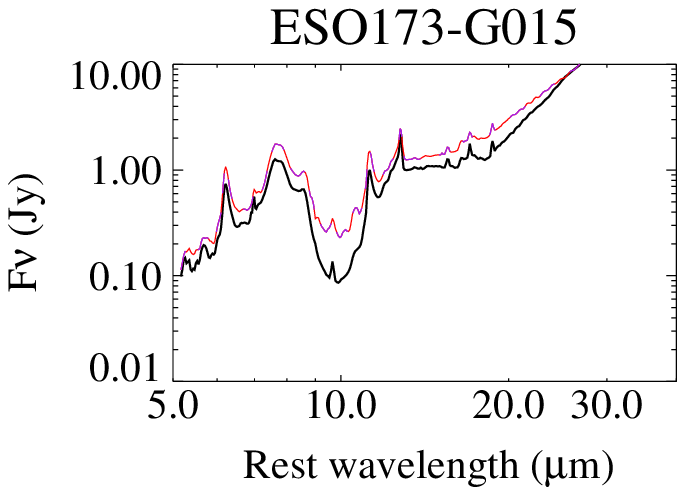}
\includegraphics[width=0.32\textwidth]{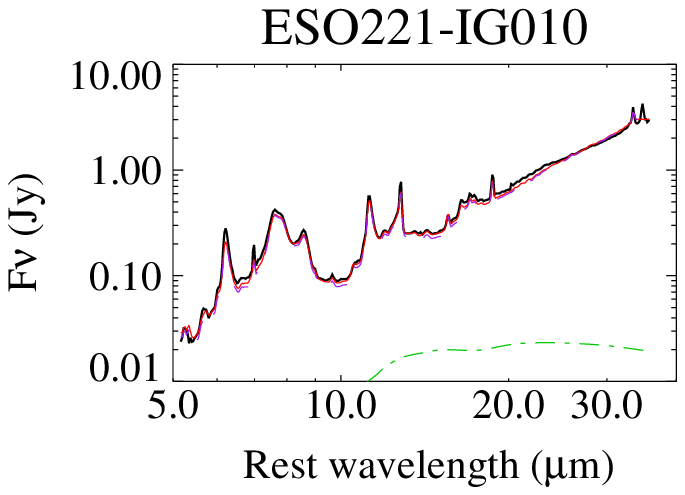}

\includegraphics[width=0.32\textwidth]{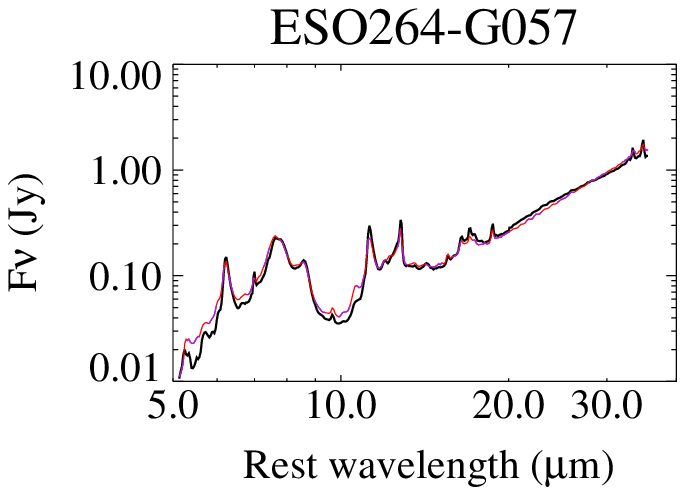}
\includegraphics[width=0.32\textwidth]{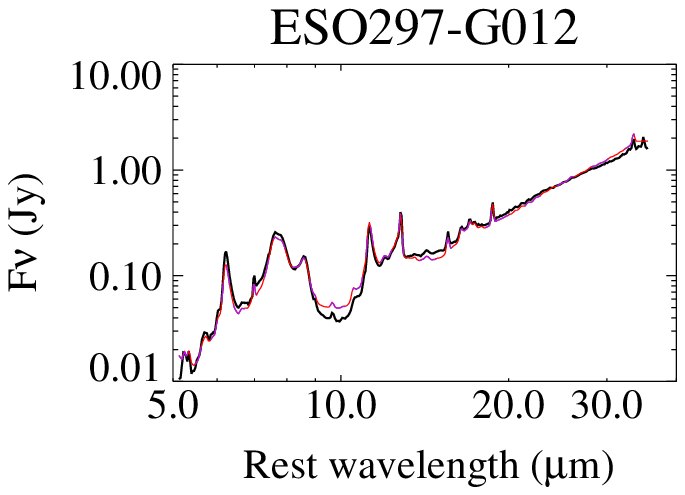}
\includegraphics[width=0.32\textwidth]{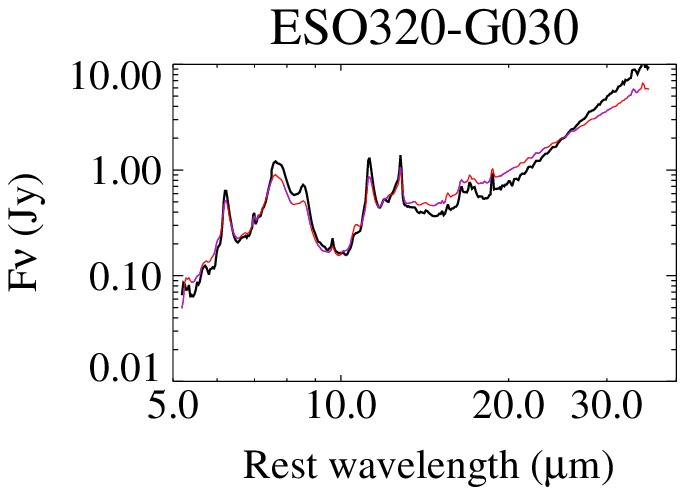}

{{\bf Figure~A2}. AGN+SB decomposition fits for those LIRGs without a mid-IR
  AGN detection. Lines are as in Figure~1.}
\end{figure*}

\begin{figure*}

\includegraphics[width=0.32\textwidth]{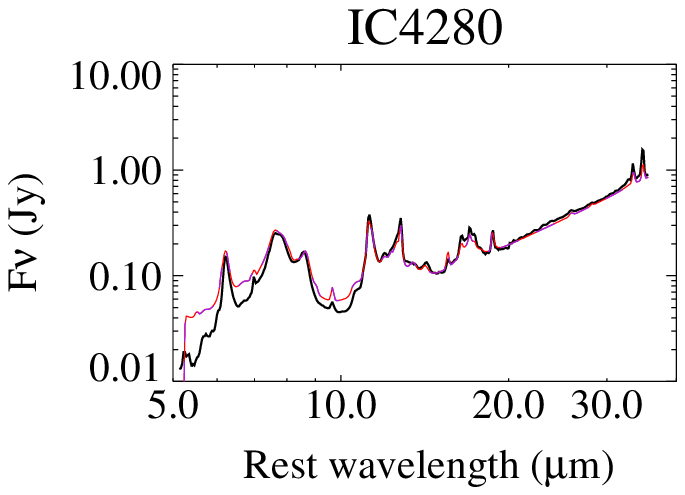}
\includegraphics[width=0.32\textwidth]{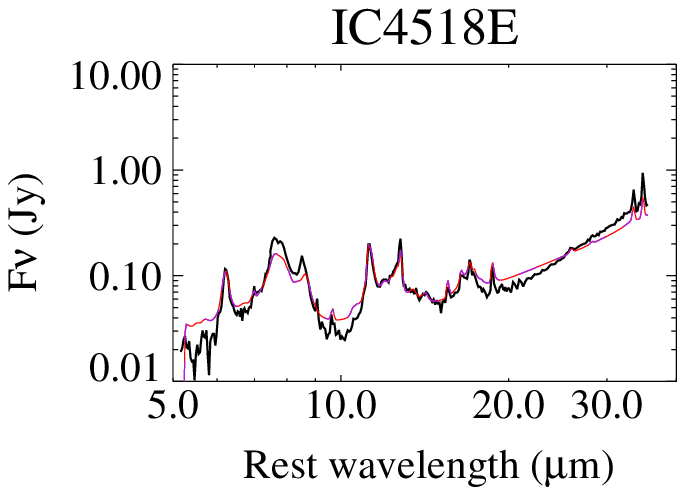}
\includegraphics[width=0.32\textwidth]{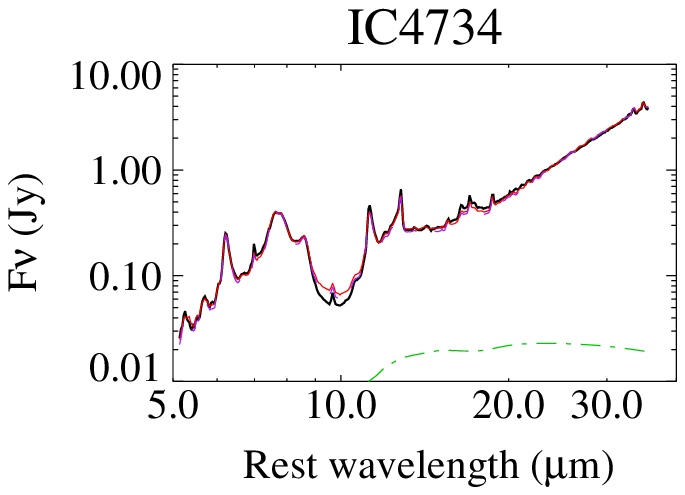}

\includegraphics[width=0.32\textwidth]{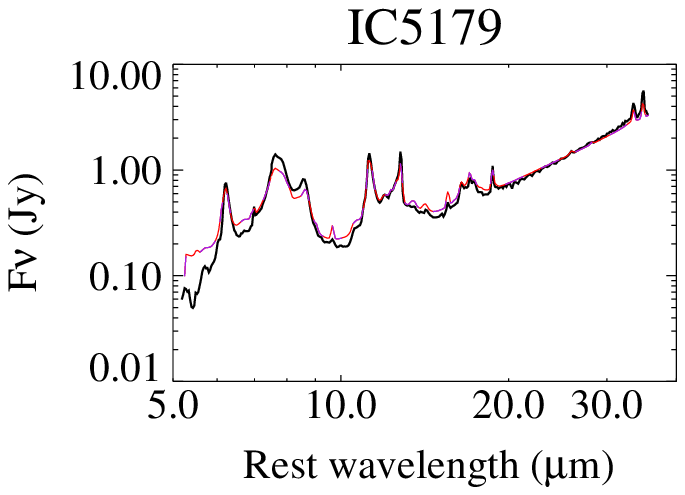}
\includegraphics[width=0.32\textwidth]{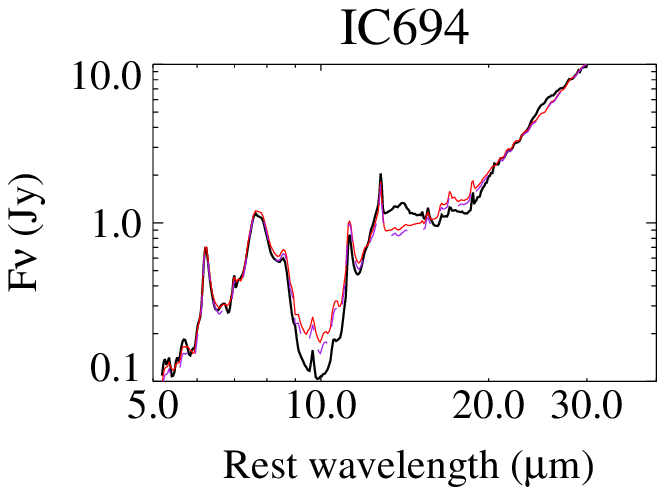}
\includegraphics[width=0.32\textwidth]{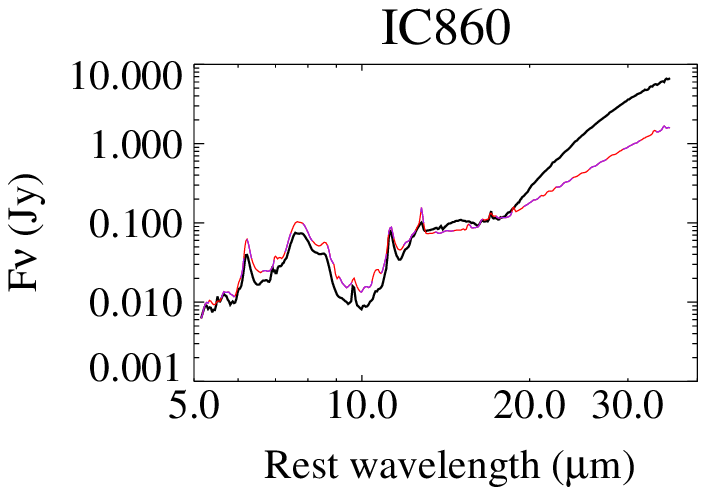}

\includegraphics[width=0.32\textwidth]{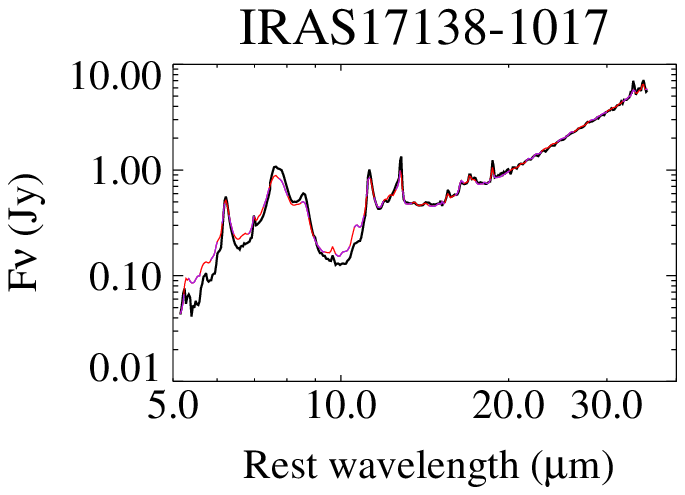}
\includegraphics[width=0.32\textwidth]{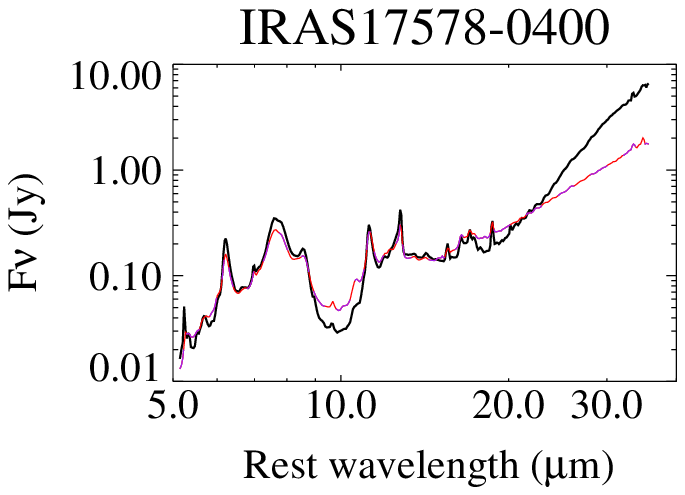}
\includegraphics[width=0.32\textwidth]{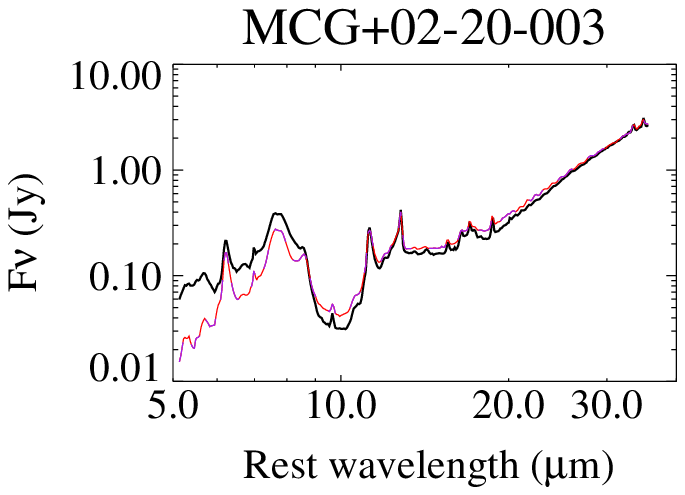}

\includegraphics[width=0.32\textwidth]{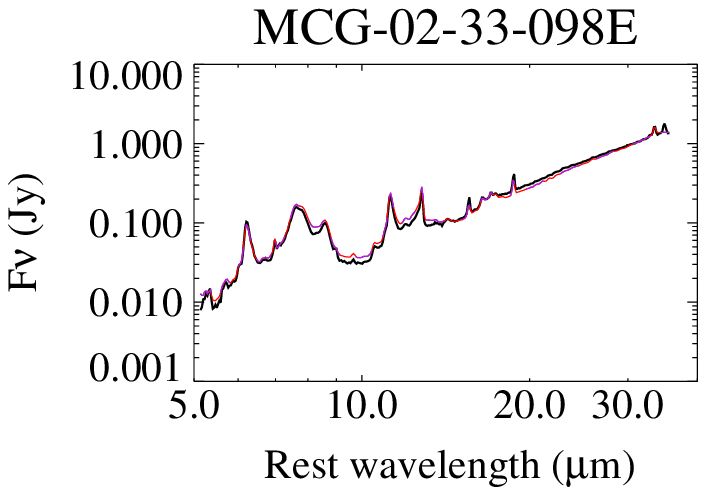}
\includegraphics[width=0.32\textwidth]{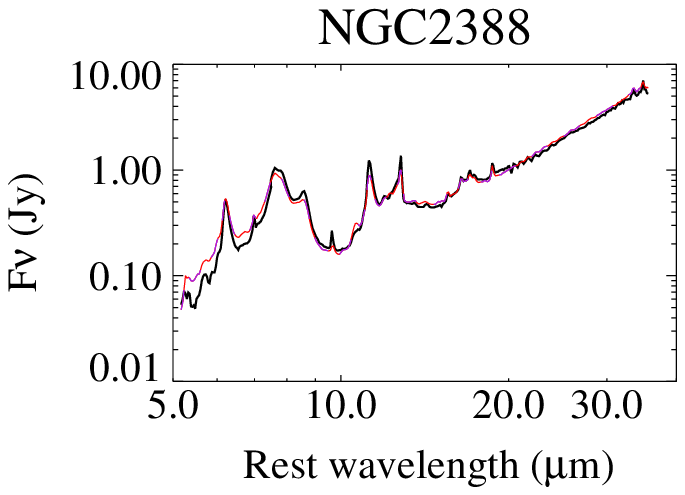}
\includegraphics[width=0.32\textwidth]{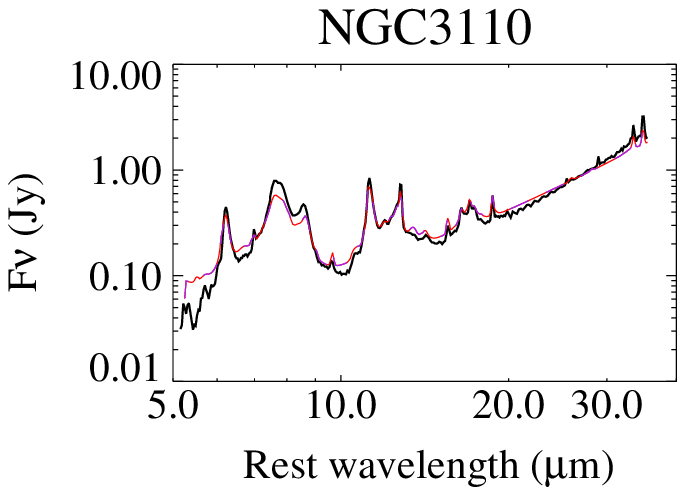}

\includegraphics[width=0.32\textwidth]{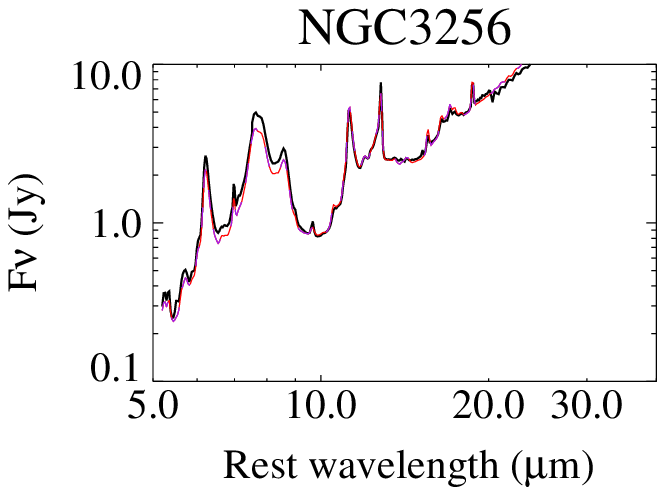}
\includegraphics[width=0.32\textwidth]{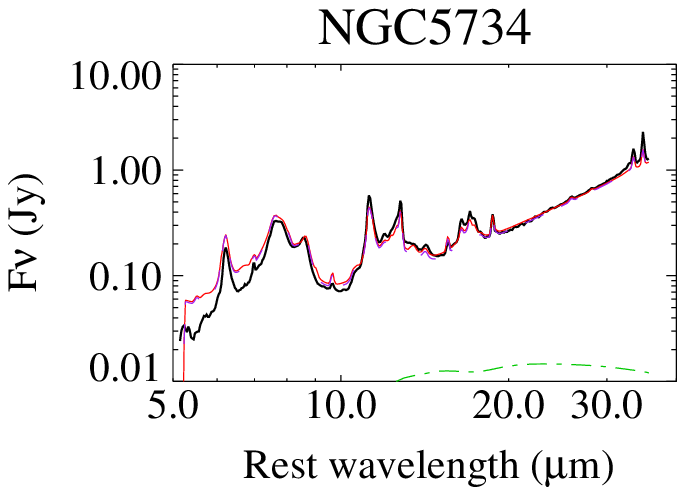}
\includegraphics[width=0.32\textwidth]{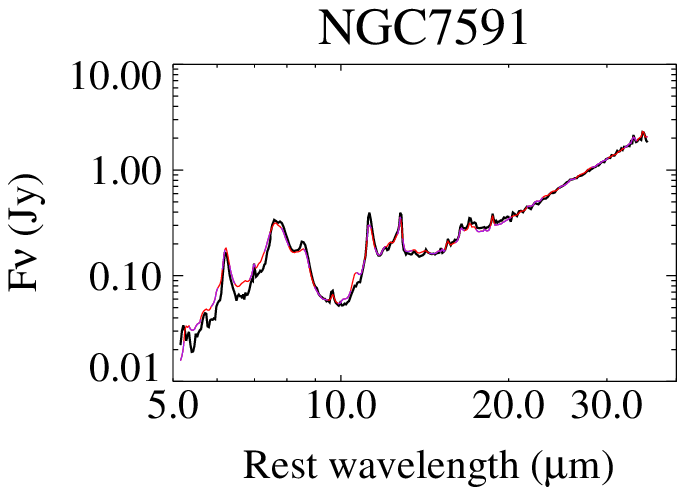}

\includegraphics[width=0.32\textwidth]{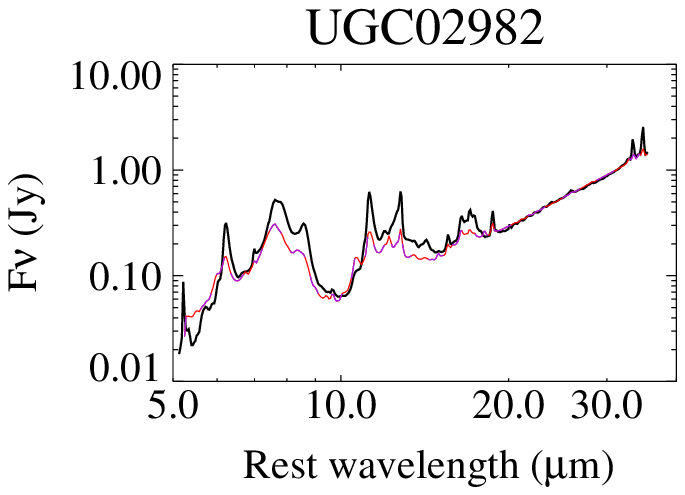}
\includegraphics[width=0.32\textwidth]{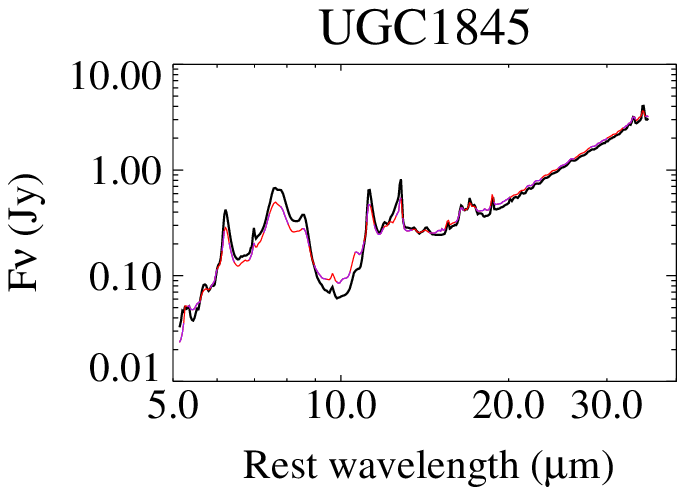}
\includegraphics[width=0.32\textwidth]{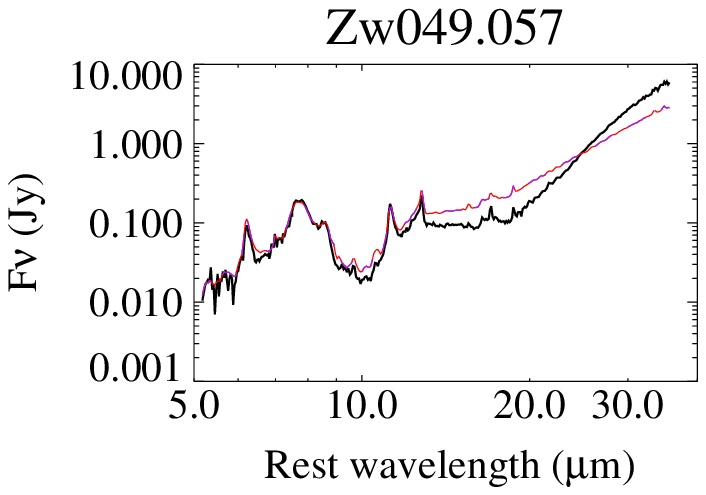}

{{\bf Figure~A2}. Continued.}
\end{figure*}

%\newpage

\end{document}